\documentclass[prl,a4paper,twocolumn,superscriptaddress,floatfix]{revtex4-1}
\usepackage[T1]{fontenc}
\usepackage{amsmath,amssymb,graphicx,graphics,rotating}
\usepackage[american]{babel}
\usepackage{dcolumn}
\usepackage{bm}
\usepackage{hyperref,color,soul,mathbbol}

\usepackage{mciteplus}

\newcommand{\ud}{\mathrm{d}}
\newcommand{\ic}{\mathrm{i}}

\newcommand{\zZ}{\mathbb{Z}}

\newcommand{\D}[1]{\displaystyle}

\begin{document} 
\title{Universal correlations in
chaotic many-body quantum states: \\ Fock-space formulation of Berry's random wave model}

\author{Florian Schoeppl}
\affiliation{Institut f\"ur Theoretische Physik, Universit\"at Regensburg, 93040 Regensburg, Germany}
\author{R\'emy~Dubertrand}
\affiliation{Department of Mathematics, Physics and Electrical Engineering, Northumbria University, NE1 8ST Newcastle upon Tyne, United Kingdom}
\affiliation{Institut f\"ur Theoretische Physik, Universit\"at Regensburg, 93040 Regensburg, Germany}
\author{Juan-Diego~Urbina}
\affiliation{Institut f\"ur Theoretische Physik, Universit\"at Regensburg, 93040 Regensburg, Germany}
\author{Klaus~Richter}
\affiliation{Institut f\"ur Theoretische Physik, Universit\"at Regensburg, 93040 Regensburg, Germany}

\date{\today}%
\begin{abstract}
The apparent randomness of chaotic eigenstates in interacting quantum systems hides subtle correlations
dynamically imposed by their finite energy per particle. These correlations are revealed when Berry’s
approach for chaotic eigenfunctions in single-particle systems is lifted into many-body space. We achieve
this by a many-body semiclassical analysis, appropriate for the mesoscopic regime of a large but finite
number of particles. We thereby identify as signatures of chaotic many-body eigenstates the universality of
both their cross-correlations and the Gaussian distribution of expansion coefficients. Combined, these two
features imprint characteristic features to the morphology of eigenstates that we check against extensive
quantum simulations. The universality of eigenstate correlations for fixed energy density is hence a further
signature of many-body quantum chaos that, while consistent with the eigenstate thermalization
hypothesis, lies beyond random matrix theory.\\
\\
\noindent DOI: \hyperlink{https://journals.aps.org/prl/abstract/10.1103/PhysRevLett.134.010404}{\color{blue}{10.1103/PhysRevLett.134.010404}}
\end{abstract}

\maketitle

Quantum systems of particles with classically chaotic
limits display universality in the statistical properties of
both their spectral fluctuations and the (spatial) correlations
of their eigenstates. This is one of the celebrated and
powerful results of semiclassical analysis, namely the study
of quantum mechanical properties in the asymptotic limit $\hbar \to 0$ pioneered by Van Vleck, Gutzwiller, and Berry,
among many others \cite{Gut1,*Gut2,Berr1,BB1970,*BB1972,*BB1974,Gutb}.


\begin{figure}[ttt]
    
    \includegraphics[width=0.95\linewidth]{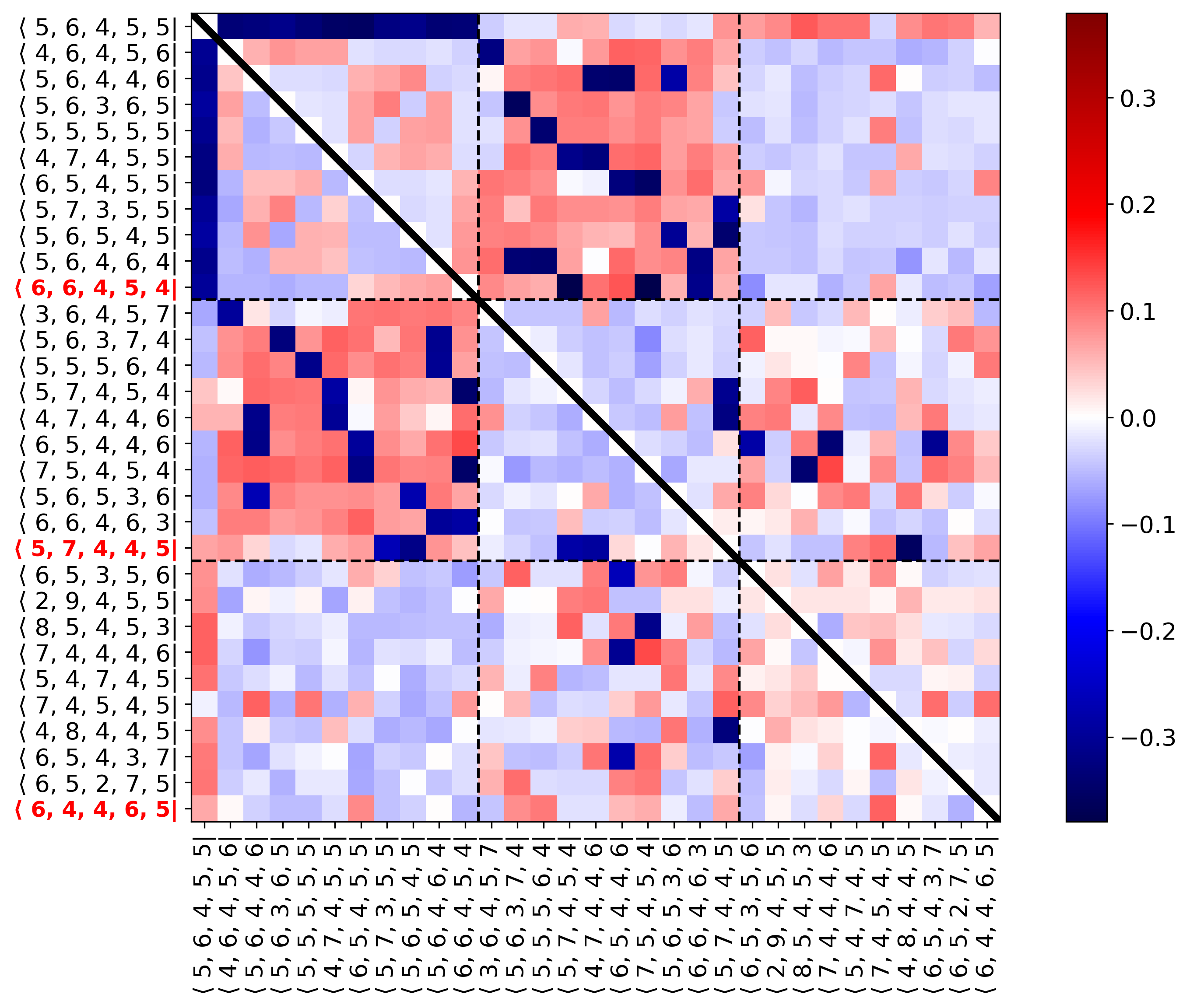}
   \caption{
   {\bf Eigenstate correlations in a Bose-Hubbard chain beyond Random Matrix Theory--}
   Cross correlations $R_{{\bf n},{\bf m}}(E)=\overline{\langle {\bf n} |\psi\rangle \langle \psi |{\bf m} \rangle}_{E}$ between expansion coefficients  (w.r.t  Fock basis states $|\mathbf n\rangle$ and $|\mathbf m\rangle$) of many-body eigenstates $|\psi^{(j)}\rangle$ of a chaotic Bose-Hubbard chain, Eq.~(\ref{eq:BH}).
   In the lower triangle results are shown for the normalized correlator $R_{{\bf n},{\bf m}}(E)/\sqrt{R_{{\bf n},{\bf n}}(E)R_{{\bf m},{\bf m}}(E)}$, obtained by averaging numerically calculated eigenstates with energies $E^{(j)}$ over a microcanonical energy window  $E^{(j)} \in [E-\eta/2,E+\eta/2]$,
   Eq.~(\ref{eq:Rnm}). They show clear mesoscopic correlations beyond random matrix theory {\em i.e.} $R^{{\rm RMT}}_{{\bf n},{\bf m}}\sim \delta_{{\bf n},{\bf m}}$, signaling quantum interference, well described by the corresponding semiclassical correlator $R^{{\rm sc}}_{{\bf n},{\bf m}}$, Eq.~(\ref{eq:Rsc}), shown in the upper triangle. For visualization, the representative states $|\mathbf n\rangle$ and $|\mathbf m\rangle$  for $L\!=\!5$ sites and $N\!=\!25$ particles
  are obtained from a reference  "seed" $|5,6,4,5,5\rangle$ by changes in local occupations up to $\pm 3$; see text and \cite{suppmat} for other seed states.
   }
	\label{fig:Corr}
\end{figure}


In the case of the single-particle (SP) “spectral statistics” ~\footnote{
Following standard terminology we use ''single particle'' instead of the more precise ''first-quantized'' nomenclature for systems where the classical limit corresponds to particle-like (instead of field-like) degrees of freedom}, this universality was embodied and linked to classical chaos in the Bohigas-Giannoni-Schmit conjecture \cite{BGS} and later confirmed by means of semiclassical theory~\cite{Berr3,Klaus1,Haak1,*Haak2}. For many-body (MB) systems the analysis of energy spectra using random matrix theory (RMT) had started in nuclear physics \cite{RevModPhys.53.385} and was later followed in condensed matter, (see~\cite{montambaux_quantum_1993}), mesoscopic physics \cite{RevModPhys.69.731} and many other domains; see the review~\cite{GUHR1998189} for an exhaustive bibliography for this early period. The study of spectral correlations has later been extended to MB systems with chaotic mean-field limit~\cite{MBQC1,Kollath_2010,PhysRevE.81.036206,Richter22}, and to systems without a semiclassical regime, see Refs. \cite{10.1143/PTPS.139.191,Rig1,PhysRevE.102.062144,PhysRevLett.121.264101,PhysRevB.91.081103,Kos2018} for a review.

For SP “eigenstate statistics” and confirmed by the
seminal numerical study \cite{mcdonaldSpectrumEigenfunctionsHamiltonian1979,mcdonaldWaveChaosStadium1988}, 
Berry's semiclassical analysis of chaotic eigenfunctions \cite{Berr4} in $d$-dimensions leads to an amplitude-amplitude correlator between positions ${\bf r}$ and ${\bf r'}$ given by the celebrated Bessel function $J_{d/2-1}(x)/x^{d/2-1}$. It displays characteristic coherent oscillations as a function of the spatial distance $x=k(E,\bar{{\bf r}})|{\bf r}-{\bf r'}|$ fixed through the local wavenumber $k^{2}(E,\bar{\bf r})=2m(E-V(\bar{\bf r}))/\hbar^{2}$ at the energy $E$ and mean position $\bar{{\bf r}}=({\bf r}+{\bf r'})/2$. Berry's result is universal in the sense of having the exact same dependence on the external potential $V({\bf r})$ for any chaotic system, an universality that moreover lies beyond  RMT for any finite energy $E < \infty$ . Once Berry's correlator fixes an amplitude distribution universally given in Gaussian form for chaotic  systems, the resulting SP Random Wave Model (RWM) \cite{Berr4,Sied1,JDUre} turns into a powerful approach to describe the spatial morphology of eigenfunctions in mesoscopic chaotic systems \cite{RWM3,Urbina06,RWM2}. Its validity is supported by a huge amount of numerical \cite{PhysRevA.37.3067,RWMBe3,bogomolny2002percolation} and experimental \cite{HJ1,HJ2,HJ3} results. Moreover, due to its appealing form based on Gaussian random fields, it allows for the explicit analytical computation of even delicate topological features \cite{RWMBe3,RWMBe4,Dennis1,Dennis2} including nodal structure \cite{bogomolny2002percolation,jainNodalPortraitsQuantum2017}.

In the MB context, the description of chaotic eigenstates has a long story dating back to the 70's where 
they were characterized through statistical measures such as the entropy \cite{lubkinEntropyNsystemIts1978,pageAverageEntropySubsystem1993,senAverageEntropyQuantum1996}. In a further refinement, 
the Eigenstate Thermalization Hypothesis (ETH) 
\cite{ETH1,ETH2,ETH3,ETH4}, 
and related concepts \cite{jainQuantumChaosRandom1997} have provided tools to relate equilibrium properties to chaotic dynamics \cite{ETH2,ALONSO1996812,Gaspard1997}. Within the resurgence of MB quantum chaos during the last two decades, MB eigenstates at the universal level have been mainly described within  pure RMT approaches where they are assumed to behave as random vectors, only weakly correlated through their orthonormality \cite{Rig1}.

\begin{figure}
	\includegraphics[width=0.95\linewidth]{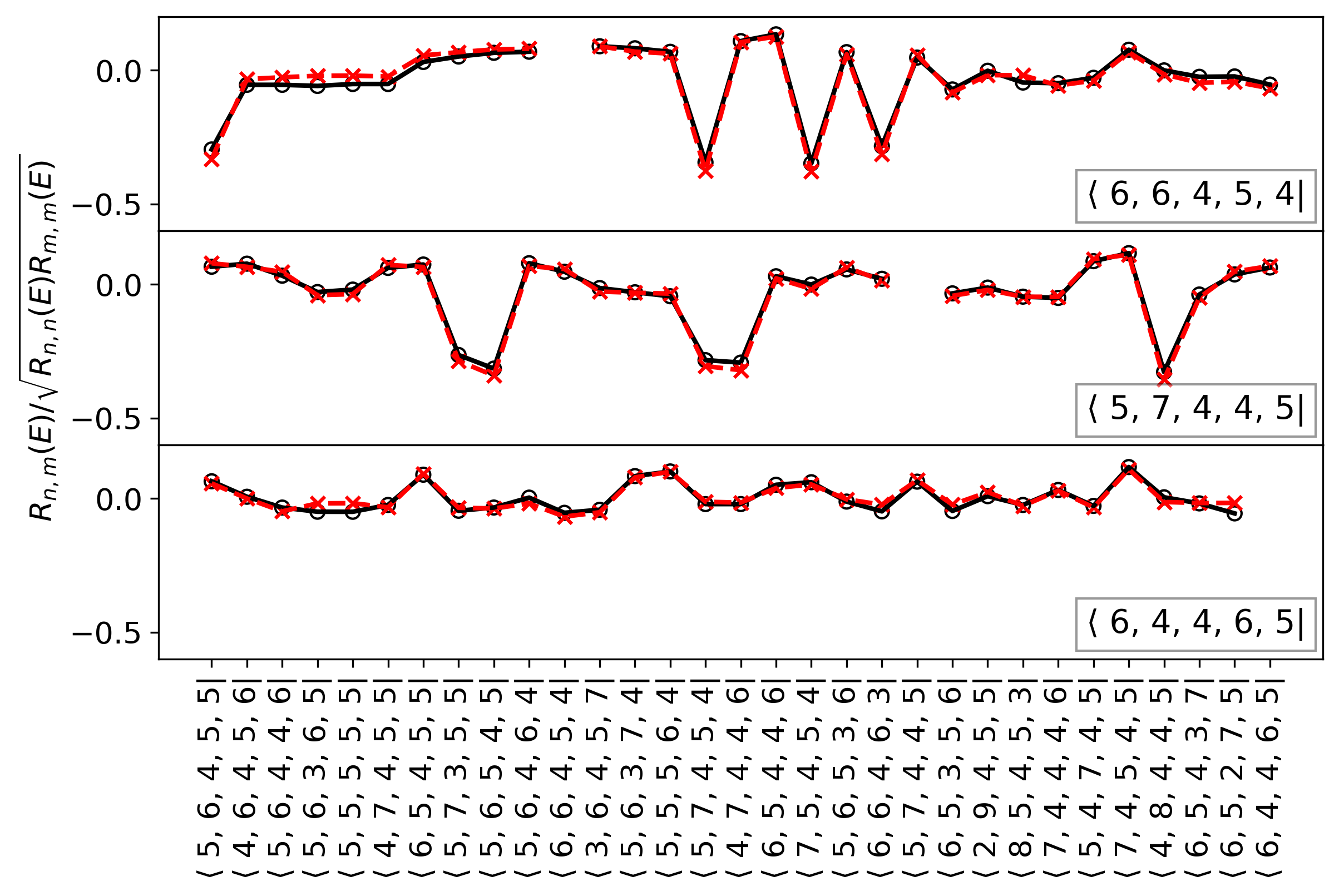}
   \caption{\textbf{Individual normalized correlators --}  The same as in Fig.~\ref{fig:Corr} but for a selection of three horizontal cuts at $|\mathbf n\rangle$ with ${\mathbf n}=6,4,4,6,5$ (bottom), ${\mathbf n}=5,7,4,4,5$ (middle), and ${\mathbf n}=6,6,4,5,4$ (top).
   Along the horizontal axis we mark a representative set of states $|\mathbf m\rangle$ with changes up to $\pm 3$ in their occupation numbers relative to the respective $|\mathbf n\rangle$.
   Black and red symbols mark numerically exact normalized correlators and the prediction (\ref{eq:Rsc}) of the MB semiclassical theory, respectively (lines serve as guides to the eye). The gaps correspond to the diagonal ${\bf n}={\bf m}$, where the normalized correlator is unity by construction.}
\label{fig:cuts}
\end{figure}

However, as indicated
in Figs.~\ref{fig:Corr} and~\ref{fig:cuts}, there exist distinct eigenstate correlations in the Fock space of chaotic (here bosonic) MB systems that are obviously not captured by RMT. Such correlations prevail, even in the ergodic phase, for systems with large but finite particle numbers, the regime where state-of-art cold-atom based quantum simulators operate~\cite{doi:10.1126/science.aal3837,Aidelsburger_2018,RevModPhys.80.885}. The swift progress of quantum state tomography \cite{Lanyon2017} appropriate for this \emph{mesoscopic} regime, demands an understanding of these correlations and their possible universality beyond RMT.

\begin{figure}[bbb]
	\includegraphics[width=0.95\linewidth]{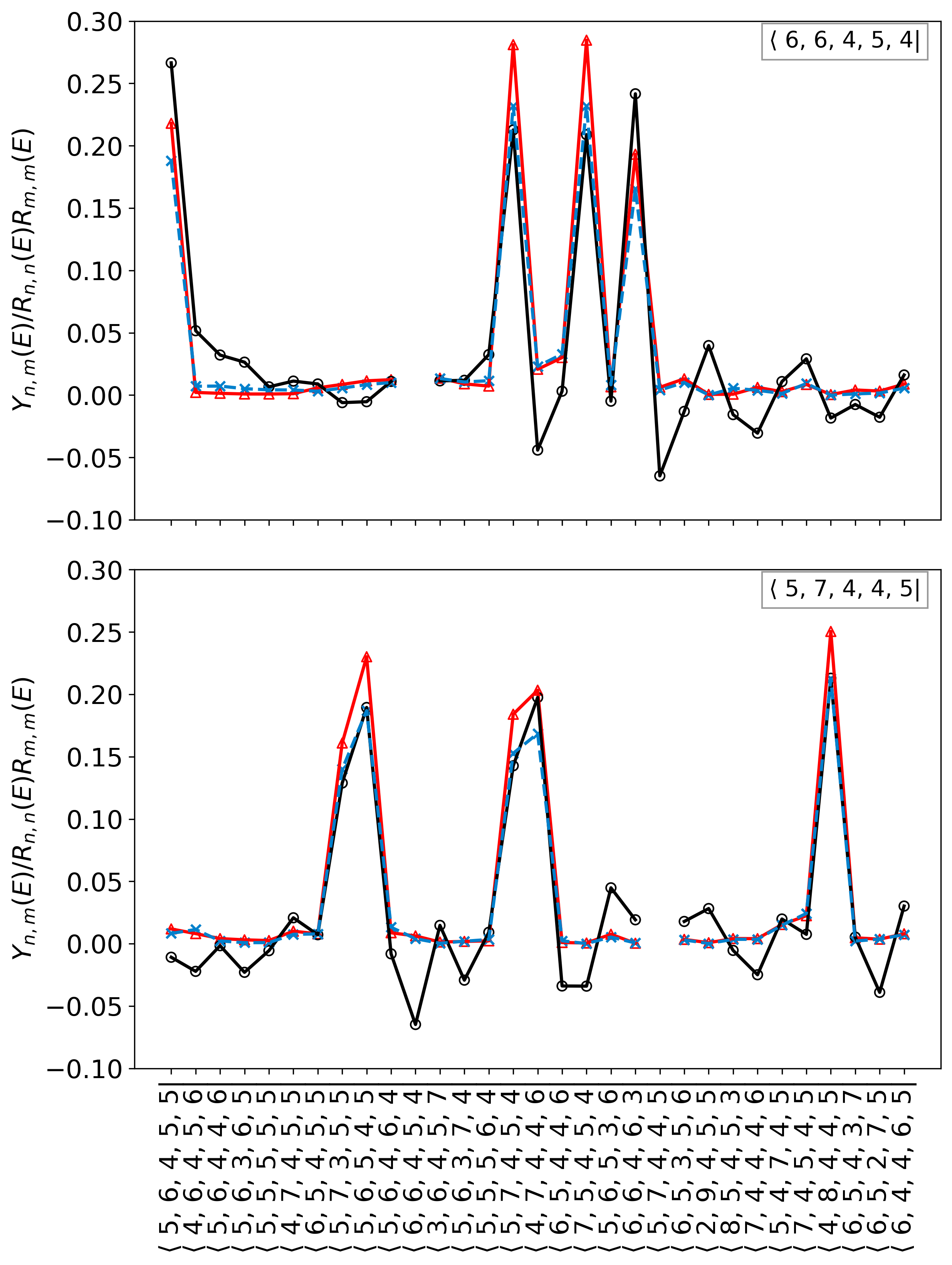}
	
   \caption{{\bf Intensity correlators --} Numerical check of the conjectured multivariate Gaussian distribution, Eq.~(\ref{eq:Gauss}), for the expansion coefficients of chaotic many-body eigenstates. Black symbols and lines: Numerically computed normalized  intensity-intensity correlator $Y_{{\mathbf n},{\mathbf m}}(E)/(R_{{\mathbf m},{\mathbf m}}(E) R_{{\mathbf n},{\mathbf n}}(E))$, defined in Eq.~(\ref{def:y_mn}) with spectral average, Eq.~(\ref{eq:Fexact}).
   Dashed blue curves: $Y_{{\mathbf n},{\mathbf m}}^{G}(E)/(R_{{\mathbf n},{\mathbf n}}(E)R_{{\mathbf m},{\mathbf m}}(E))$ with Gaussian assumption $Y_{{\mathbf n},{\mathbf m}}^{G}(E)=2|R_{{\mathbf n},{\mathbf m}}(E)|^{2}$, Eq.~(\ref{def:yg_mn}). Top and bottom panels show cross correlators between a set of Fock states $|{\bf m}\rangle$, marked along the horizontal axis, and reference states $|{\mathbf n}\rangle=|6,6,4,5,4\rangle$ and $|5,7,4,4,5\rangle$, respectively. The prediction based on the full MB random wave model with semiclassical approximation $R_{{\mathbf n},{\mathbf m}}(E)\simeq R^{\rm sc}_{{\mathbf n},{\mathbf m}}(E)$ (Eq.~(\ref{eq:Rsc}), red curve) is closely following the numerical data based on the Gaussian approximation (blue asterisks). The gaps mark  cases ${\bf n}={\bf m}$ where it coincides with $Y^{\rm RMT}_{{\mathbf n},{\mathbf m}}/R^{{\rm RMT}}_{{\bf n},{\bf n}}R^{{\rm RMT}}_{{\bf m},{\bf m}}\!=\!2\delta_{{\mathbf n},{\mathbf m}}$.}
\label{fig:CorrInt}
\end{figure}

Hence, in this Letter, we provide a generalization of the RWM and the extension of Berry's ansatz into the mesoscopic regime of chaotic MB systems. We will lift its key ingredients into the Fock space of quantum fields, namely, the presence of microscopic correlations predicted by semiclassics that lie beyond RMT, as seen in
Figs.~\ref{fig:Corr} and \ref{fig:cuts}, and through the conjectured multivariate Gaussian distribution of expansion coefficients (Fig.~\ref{fig:CorrInt}).   

To this end we need to extend quantum chaos techniques that successfully described SP systems \cite{Gutb,Haake06}, into the asymptotic regime of large but finite particle number $N\gg 1$. 
This is one of the quest of MB quantum chaos, a field that has swiftly moved at the frontier of several fundamental questions, ranging from equilibration in statistical physics~\cite{MBQC1,Rig1,MBQC3,Kos2018} to gravity \cite{JT1,JT2}. Discrete bosonic fields such as the Bose-Hubbard chains considered here possess indeed a classical limit for $N\rightarrow \infty$, amenable to a mean-field description. Then 
$1/N \ll 1$ plays the role of an effective Planck's constant and semiclassical techniques for chaotic SP systems have been generalized to MB (Fock) space, see \cite{Richter22} for a recent review.  
In this way genuine MB interference effects, ranging from MB coherent backscattering \cite{CBS} and MB spin echo \cite{MBSE}, via MB spectra from periodic mean field solutions \cite{TF_Thomas,TF_Remy} to
out-of-time-order-correlators \cite{Josef,QB} have been understood in semiclassical terms. This semiclassical approach fully accounts for MB quantum interference \cite{PhysRevA.97.061606}, beyond the (incoherent) truncated Wigner approximation, see e.g.~\cite{PhysRevA.68.053604,pappalardi2020quantum}.


To introduce the statistical measures to describe eigenstate fluctuations,
consider a  MB system defined by a Hamiltonian $\hat{H}$ with  eigenfunctions $\hat{H}|\psi^{(j)}\rangle=E^{(j)}|\psi^{(j)}\rangle$ and eigenvalues $E^{(0)}\le E^{(1)} \le \ldots$ \footnote{
If necessary, systematic degeneracies due to symmetries are excluded by focusing on the  eigenfunctions within a given symmetry-related subspace with fixed values of the corresponding quantum numbers}. Consider now a spectral region of width $\eta$ around an energy $E$, implemented through a window function $W_{\eta}(x)$ centered at $x=0$, and a functional $F(|\psi\rangle)$ associating a complex number to an arbitrary state $|\psi\rangle$. We then define
\begin{equation}
\label{eq:Fexact}
\overline{F(|\psi\rangle)}_{E}=\sum_{j=0}^{\infty}\frac{W_{\eta}(E^{(j)}\!-\!E)}{\rho_{\eta}(E)} F(|\psi^{(j)}\rangle)\, 
\end{equation}
as the average of the fluctuating numbers $F(|\psi^{j}\rangle)$ over the spectral window. Normalization $\bar{1}_{E}=1$ gives then
\begin{equation}
 \rho_{\eta}(E)={\rm Tr~}W_{\eta}(E-\hat{H}) \sim {\rm e}^{N}
\end{equation}
 as the mean ({\em i.e.}~coarse-grained) level density \footnote{We assume $\eta$ to be large compared to the local mean level spacing which implies a large number of eigenstates inside the spectral window.}. 

In bosonic MB systems, the Hilbert space is spanned by the (Fock) common eigenstates $|{\bf n}\rangle$ of the occupation number operators $\hat{n}_{\alpha}=\hat{b}_{\alpha}^{\dagger}\hat{b}_{\alpha}$ ($\alpha = 1, \ldots, L)$ where $n_\alpha$ counts the number of particles and $\hat{b}_{\alpha}$ ($\hat{b}_{\alpha}^{\dagger}$) annihilates (creates) a particle in the $\alpha$-th single-particle state. The choice $F(|\psi\rangle)=\langle {\bf n}|\psi\rangle \langle \psi|{\bf m}\rangle$ in Eq.~(\ref{eq:Fexact}) defines then the two-point cross-correlation of expansion coefficients $\langle {\bf n}|\psi^{(j)}\rangle$ and $\langle {\bf m}|\psi^{(j)}\rangle$ over the spectral window:
\begin{eqnarray}
\label{eq:Rnm}
\overline{\langle {\bf n}|\psi\rangle \langle \psi|{\bf m}\rangle}_{E}&=&\sum_{j=0}^{\infty}\frac{W_{\eta}(E^{(j)}\!-\!E)}{\rho_{\eta}(E)}\langle {\bf n}|\psi^{(j)}\rangle \langle \psi^{(j)}|{\bf m}\rangle \nonumber \\ &\equiv&R_{{\bf n},{\bf m}}(E) \; .
\end{eqnarray}
This in turn naturally defines the covariance matrix
\begin{equation}
\label{eq:CovR}
    R_{{\bf n},{\bf m}}(E)=\langle {\bf n}| \hat{R}(E)|{\bf m}\rangle,
\end{equation}
in terms of the basis-independent covariance operator
\begin{equation}
  \hat{R}(E)=\overline{|\psi\rangle\langle \psi |}_{E}=\frac{W_{\eta}(E-\hat{H})}{{\rm Tr~}W_{\eta}(E-\hat{H})} \; , 
  \label{eq:R}
\end{equation}
{\it i.e.}~the density operator corresponding to the microcanonical window $W_{\eta}(x)$. Such intrincate mesoscopic fluctuations of the covariance matrix were numerically observed in Ref.~\cite{nakerst2021eigenstate}, and one of our goals here is to provide a theoretical understanding of their structure. 

We implement the MB generalization of Berry's construction in three stages: First, it is assumed that the quantum MB system possesses a well defined chaotic classical (mean-field) limit and a semiclassical regime. Second, we employ for
$\overline{F(|\psi\rangle)}_{E}$
a Gaussian distribution, 
\begin{equation}
\label{eq:Gauss}
\overline{F(|\psi\rangle)}_{E}\!\simeq\!
\overline{F(|\psi\rangle)}_{E}^{\rm G}\!=\!
\frac{1}{\mathcal{N}(\hat{R})}\int_{{\cal H}}\!\!F(|\psi\rangle){\rm e}^{-\langle \psi |\hat{R}(E)^{-1}|\psi\rangle}\ud|\psi\rangle\ ,
\end{equation}
valid in chaotic systems~\cite{Sied1,JDUre,Heller_2007}, where $\mathcal{N}(\hat{R})$ fixes $\overline{1}_{E}^{\rm G}=1$. At a third stage, we use a semiclassical approximation for the correlation matrix $\hat{R}(E)$ instead of its exact and usually untractable microscopic expression \footnote{Knowledge of $\hat{R}(E)$ provides all information required for the calculation of the microcanonical (mc) expectation value of any MB observable $\hat{O}$, as $\langle \hat{O} \rangle^{(\rm mc)}_{E}={\rm Tr}\left[\hat{R}(E)\hat{O}\right]$. Its exact determination is therefore a formidable task in general}, where both universal and system-specific aspects enter.

To be definite, we now implement these steps and construct the MB RWM for systems described by Bose-Hubbard-type Hamiltonians
\begin{equation}
\hat{H}=\!\sum_{\alpha=1}^{L} \!\!
\left[\!-t(\hat{b}^{\dagger}_{\alpha}\hat{b}_{\alpha+1}\!+\!\hat{b}^{\dagger}_{\alpha+1}\hat{b}_{\alpha})\!+\!\epsilon_{\alpha}\hat{n}_{\alpha}\!+\!U\hat{n}_{\alpha} (\hat{n}_{\alpha}\!-\!1) \right]
\label{eq:BH}
\end{equation}
where the on-site energies $\epsilon_\alpha$ are chosen in a way to break any discrete symmetry, and periodic boundary conditions are assumed. This model describes a wide range of physical systems, see e.g. \cite{PhysRevB.40.546,krutitsky2016ultracold,bloch2008many} and references therein. It has a classical limit that corresponds to large occupations $n_{\alpha} \gg 1$ with a classical Hamiltonian $H_{\rm cl}({\boldsymbol \rho},{\boldsymbol \theta})$ obtained from $\hat{H}$ by the replacements $\hat{b}_\alpha \to \sqrt{\rho_\alpha}{\rm e}^{-i\theta_\alpha},\hat{b}_\alpha^{\dagger} \to \sqrt{\rho_\alpha}{\rm e}^{i\theta_\alpha}$ \cite{TF_Thomas,TF_Remy}. 
The classical phase space, endowed with the canonical variables $({\boldsymbol \rho},{\boldsymbol \theta})$,  was numerically found~\cite{TF_Remy,Steve} to be  integrable in the limiting regimes $t/UN\to 0, \infty$, while chaotic dynamics sets in for $t/UN\sim 1$.


To proceed with the MB semiclassical approximation we consider the Wigner transform  of $\hat{R}(E)$, Eq.~(\ref{eq:R}), \footnote{We do not address the issue of defining the Wigner transform on discrete spaces, as it is irrelevant in the semiclassical regime considered here.}
\begin{equation}
\label{eq:Wigner}
    {\cal W}_{E}({\boldsymbol \rho},{\boldsymbol \theta})=\sum_{{\bf m} < {\boldsymbol \rho}} R_{{\boldsymbol \rho}+{\bf m},{\boldsymbol \rho}-{\bf m}}(E) \, {\rm e}^{i{\boldsymbol \theta}\cdot {\bf m}} \,  ,
\end{equation}
where the notation ${\bf m} < {\boldsymbol \rho}$ means $m_{\alpha}<\rho_{\alpha}$ for all $\alpha$. We invoke now the Fock space version of the eigenstate condensation hypothesis of Voros and Berry~\cite{voros1976semi,berry1977semi} in the form ${\cal W}_{E}\simeq {\cal W}_{E}^{{\rm cl}}$, with classical phase space distribution
\begin{equation}
\label{eq:RW}
{\cal W}_{E}^{{\rm cl}}({\boldsymbol \rho},{\boldsymbol \theta})=\frac{1}{\rho_{\eta}^{{\rm cl}}(E)}W_{\eta}\left(E\!-\!H_{\rm cl}({\boldsymbol \rho},{\boldsymbol \theta})\right) \, .
\end{equation}

The level density $
\rho_{\eta}^{{\rm cl}}(E)=\int W_{\eta}\left(E\!-\!H_{\rm cl}({\boldsymbol \rho},{\boldsymbol \theta})\right) d{\boldsymbol \rho}d{\boldsymbol \theta}$ is
 defined by the volume of the classical energy shell of width $\eta$. Inverting relation~(\ref{eq:Wigner}) to obtain the corresponding semiclassical form of the covariance matrix $R^{\rm sc}$ is in general a formidable task, but thanks to the local character of $\hat{H}$ we get as our main result (see \footnote{See Supplemental Material at [url], which further includes \cite{Fischer_2013} for additional information about the derivation of the Eq.(\ref{eq:Rsc}) and \cite{beugeling2018statistical} for an in depth discussion about statistical properties of eigenstate amplitudes in complex quantum systems.} for details)
\begin{widetext}
\begin{equation}
\label{eq:Rsc}
\hspace{-0.3cm}R^{{\rm sc}}_{{\bf n},{\bf m}}(E)=\frac{1}{\rho_{\eta}^{{\rm cl}}(E)}\sum_{Q\in\zZ}  \int_{-\infty}^\infty\frac{\ud \tau}{2\pi} \tilde{W}(\tau)\exp{\{\ic \tau [E\!-\! U \sum_{\alpha} I_\alpha(I_\alpha\!-\!1) \!-\! \sum_{\alpha} \varepsilon_\alpha I_\alpha]}\} \prod_{\alpha=1}^L
  e^{\ic \frac{\pi}{2}(\delta_{\alpha}+ Q)} J_{\delta_{\alpha}+ Q}(2t\tau\sqrt{I_\alpha I_{\alpha+1}}) \, .
\end{equation}
\end{widetext}
Here $\tilde{W}(\tau)$ is the Fourier transform of $W_{\eta}(E)$,
$J_{l}(x)$ the $l$th Bessel function, ${\bf I}= ({\bf n}+{\bf m})/2$ with $I_{L+1}=I_{1}$, and $\delta_\alpha = \sum_{\beta=1}^\alpha (n_\beta - m_\beta)$.
    
Together with Eq.~(\ref{eq:Gauss}) the  expression~(\ref{eq:Rsc}) allows us to compute the average of \textit{any} functional $F$ in Eq.~(\ref{eq:Fexact}) within our MB RWM. It is therefore important to compare this central analytical prediction with corresponding numerical results for the Bose Hubbard model, Eq.~(\ref{eq:BH}). For definiteness we choose $L=5,N=25$ and $UN=2t$, where the classical dynamics is predominantly chaotic. In Figs.~\ref{fig:Corr} and~\ref{fig:cuts} we compare the normalized correlators $R_{{\bf n},{\bf m}}(E)/\sqrt{R_{{\bf n},{\bf n}}(E)R_{{\bf m},{\bf m}}(E)}$ obtained by exact diagonalization and averaged over a spectral window comprising $\sim 10^{3}$ MB states, with the MB semiclassical approximation (\ref{eq:Rsc}). In Fig.~(\ref{fig:cuts}) we show the correlations between reference Fock states  $|\mathbf n\rangle =|6,6,4,5,4\rangle, |5,7,4,4,5\rangle,|6,4,4,6,5\rangle$ and representative states $|\mathbf m\rangle$ indicated along the horizontal axis. In all three cases we find remarkable agreement between the semiclassical prediction (red symbols) and the numerical results (black symbols). Instead of fixing a single reference basis state $|\mathbf n\rangle$,  the color matrix in Fig. \ref{fig:Corr} represents cross correlations among all Fock basis states around $|5,6,4,5,5\rangle$. The upper right and lower left triangle shows the semiclassical and the numerical results, respectively. The apparent symmetry, quantitatively confirmed by the cuts shown in Fig.~(\ref{fig:cuts}), again reflects the excellent agreement, see also \cite{suppmat} for further comparisons. 


After this exhaustive check of the semiclassical approximation for the covariance matrix we focus on the second essential aspect of the MB RWM, namely the underlying Gaussian assumption for the joint distribution of eigenstate amplitudes, encoded in Eq.~(\ref{eq:Gauss}). This assumption has far reaching consequences well beyond the well-known Gaussian distribution of expansion coefficients along {\it a single fixed} $|{\mathbf n}\rangle$ reported in \cite{atas2017quantum,nakerst2021eigenstate} and numerically confirmed \cite{suppmat} for chaotic Bose-Hubbard models. In fact, Eq.~(\ref{eq:Gauss}) describes the {\it joint} probability distribution of {\it the whole set} of coefficients $\langle {\mathbf n}|\psi^{(j)}\rangle$ for {\it all} $|{\bf n}\rangle$. A prominent statistical measure, sensitive to both the microscopic correlations and the multivariate Gaussian nature of the  expansion coefficients, is the Fock space version of the two-point intensity correlator~\cite{MIRLIN2000},
\begin{equation}
Y_{{\mathbf n},{\mathbf m}}(E)=\overline{|\langle {\mathbf n}|\psi\rangle|^{2}|\langle {\mathbf m}|\psi\rangle|^{2}}_{E}-\overline{|\langle {\mathbf n}|\psi\rangle|^{2}}_{E} \overline{|\langle {\mathbf m}|\psi\rangle|^{2}}_{E}  \, .
\label{def:y_mn}
\end{equation}
Using the properties of multivariate Gaussian distributions in Eq.~(\ref{eq:Gauss}), we easily get \cite{suppmat}
\begin{equation}
Y_{{\mathbf n},{\mathbf m}}^{G}(E)=2|R_{{\mathbf n},{\mathbf m}}(E)|^{2} \label{def:yg_mn}
\end{equation}
for the case of real expansion coefficients considered here. In Fig.~\ref{fig:CorrInt} we check this prediction by performing  numerical simulations independently for both 
$Y_{{\mathbf n},{\mathbf m}}(E)$ (black curves) and $|R_{{\mathbf n},{\mathbf m}}(E)|^2$ (dashed blue) for two representative seed states \footnote{Similar results are obtained for any seed state under the conditions discussed below}. We see that mesoscopic MB eigenstates behave as realizations of a {\it correlated} multivariate Gaussian distribution, up to deviations expected for such higher-order correlators due to stronger finite-size effects.

Combining the Gaussian conjecture with the semiclassical approximation $R_{{\mathbf n},{\mathbf m}}(E)\simeq R^{\rm sc}_{{\mathbf n},{\mathbf m}}(E)$ yields moreover closed analytical results for
$Y_{{\mathbf n},{\mathbf m}}^{G}(E)$, 
Eq.~(\ref{def:yg_mn}). In Fig.~(\ref{fig:CorrInt}) they are 
indicated as red asterisks again coinciding well with numerical Gaussian theory (blue).
This close agreement gives full support of the proposed MB version of the RWM. The semiclassical covariance matrix captures the \emph{microscopic} correlations of typical eigenstates responsible for the structure of any correlation function. Thus, the semiclassical correlator unveils MB information well beyond RMT that predicts energy-independent flat profiles as $Y^{\rm RMT}_{{\mathbf n},{\mathbf m}}/R^{{\rm RMT}}_{{\bf n},{\bf n}}R^{{\rm RMT}}_{{\bf m},{\bf m}}\!=\!2\delta_{{\mathbf n},{\mathbf m}}$.

We briefly comment now on some important instances where our analysis touches previous results and concepts.

{\it On the notion of universality.} As mentioned before, the universality of Berry's RWM stems from the universal functional dependence of the two-point amplitude correlator on the external potential as long as, i) it generates strongly chaotic dynamics (nonzero Lyapunov exponent) and, ii) the kinetic term is quadratic in the momentum. In the same venue, as long as the hopping term of $\hat{H}$ is nearest-neighbor only, our results will hold for \textit{any} interaction term $H^{\rm int}(\hat{\mathbf n})$ that is diagonal in the Fock basis and generates chaotic mean-field dynamics, the only change being the substitution 
\begin{equation}
U\sum_{\alpha}I_{\alpha}(I_{\alpha}-1) \to H^{\rm int}({\bf I})
\end{equation}
in the semiclasical cross-correlator, Eq.~(\ref{eq:Rsc}).

{\it Relation with ETH and Behemoths.} The predictions of the MB RWM  are consistent with the ETH, as expectation values $\langle \hat{O} \rangle ={\rm Tr~}[\hat{O}\hat{R}]$ of MB observables $\hat{O}$ take the ETH form in leading order through Eq.~(\ref{eq:RW}). Our results extend the usual ergodic arguments for diagonal matrix elements into a theory for cross-correlations of expansion coefficients, and this is achieved by means of the same object, namely the microcanonical covariance matrix. Despite their common roots, in general the RWM and the ETH address different objects, {\em i.e.}, Fock-space correlations and expectation values. There exists however a type of non-local MB observables, the so-called Behemoths introduced in \cite{Haque2019} like $\hat{\Omega}_{{\bf n},{\bf m}}=|{\bf n}\rangle \langle{\bf m}|+|{\bf m}\rangle \langle{\bf n}|$, whose microcanonical expectation values are {\it exactly} given by $R_{{\mathbf n},{\mathbf m}}(E)$. Our analysis implies that Behemoths display universal structure at fixed energy, beyond the fidelity to RMT in the infinite temperature limit reported in \cite{Haque2019}.

{\it Regime of validity.} Within the semiclasical regime $N\gg 1$, and assuming that the mean-field dynamics are chaotic, the key steps in Eqs.~(\ref{eq:Wigner}) and~(\ref{eq:RW}) treat the interacting term of the Hamiltonian as a constant background evaluated at the mean occupations ${\bf I}= ({\bf n}+{\bf m})/2$. This means the kinetic (hopping) contribution to the energy $E$ must be sufficiently large and, vice versa, 
that the interaction term of the classical Hamiltonian fulfills, for some constant $c$,
\begin{equation}
 \left|E-H^{\rm int}({\boldsymbol I})\right|>c.
\end{equation}
 This is confirmed by an extensive check of the validity of the MB RWM predictions across the whole Fock space~\cite{suppmat}. 

{\it Conclusions}. We have lifted Berry's random wave model from configuration space into the Fock space of interacting systems by constructing an accurate semiclassical theory describing the statistical backbone of chaotic states in mesoscopic MB systems. This MB random state model relies on the identification of chaotic eigenstates as correlated Gaussian fields arising from massive, genuine MB interference. Its central building block is a covariance matrix constructed by means of MB semiclassics techniques where the large-$N$ mean-field regime plays the role of a classical limit. 
The predictions of the MB random state model for both the multivariate and correlated features of chaotic MB eigenstates
show quantitative agreement with numerical results from 
extensive simulations for Bose-Hubbard systems. 
In this way a pillar of the field of single-particle quantum chaos, the universality of spatial fluctuations of chaotic eigenstates, is also established in the realm of many-body quantum chaos.

{\em Acknowledgements --} We thank Arnd B\"acker, Eugene Bogomolny,  Silvia Pappalardi and Mark Srednicki for valuable discussions. We acknowledge financial support from the Deutsche Forschungsgemeinschaft (German Research Foundation) through Ri681/14-1, and through Ri681/15-1 within the Reinhart-Koselleck Programme.

\bibliographystyle{apsrevM} 
\bibliography{refRW}

\ifx\mcitethebibliography\mciteundefinedmacro
\PackageError{apsrevM.bst}{mciteplus.sty has not been loaded}
{This bibstyle requires the use of the mciteplus package.}\fi
\begin{mcitethebibliography}{92}
\expandafter\ifx\csname natexlab\endcsname\relax\def\natexlab#1{#1}\fi
\expandafter\ifx\csname bibnamefont\endcsname\relax
  \def\bibnamefont#1{#1}\fi
\expandafter\ifx\csname bibfnamefont\endcsname\relax
  \def\bibfnamefont#1{#1}\fi
\expandafter\ifx\csname citenamefont\endcsname\relax
  \def\citenamefont#1{#1}\fi
\expandafter\ifx\csname url\endcsname\relax
  \def\url#1{\texttt{#1}}\fi
\expandafter\ifx\csname urlprefix\endcsname\relax\def\urlprefix{URL }\fi
\providecommand{\bibinfo}[2]{#2}
\providecommand{\eprint}[2][]{\url{#2}}

\bibitem[{\citenamefont{Gutzwiller}(1967)}]{Gut1}
\bibinfo{author}{\bibfnamefont{M.~C.} \bibnamefont{Gutzwiller}}, \bibinfo{journal}{Journal of Mathematical Physics} \textbf{\bibinfo{volume}{8}}, \bibinfo{pages}{1979} (\bibinfo{year}{1967})\relax
\mciteBstWouldAddEndPuncttrue
\mciteSetBstMidEndSepPunct{\mcitedefaultmidpunct}
{\mcitedefaultendpunct}{\mcitedefaultseppunct}\relax
\EndOfBibitem
\bibitem[{\citenamefont{Gutzwiller}(1969)}]{Gut2}
\bibinfo{author}{\bibfnamefont{M.~C.} \bibnamefont{Gutzwiller}}, \bibinfo{journal}{Journal of Mathematical Physics} \textbf{\bibinfo{volume}{10}}, \bibinfo{pages}{1004} (\bibinfo{year}{1969})\relax
\mciteBstWouldAddEndPuncttrue
\mciteSetBstMidEndSepPunct{\mcitedefaultmidpunct}
{\mcitedefaultendpunct}{\mcitedefaultseppunct}\relax
\EndOfBibitem
\bibitem[{\citenamefont{Berry and Mount}(1972)}]{Berr1}
\bibinfo{author}{\bibfnamefont{M.~V.} \bibnamefont{Berry}} \bibnamefont{and} \bibinfo{author}{\bibfnamefont{K.}~\bibnamefont{Mount}}, \bibinfo{journal}{Reports on Progress in Physics} \textbf{\bibinfo{volume}{35}}, \bibinfo{pages}{315} (\bibinfo{year}{1972})\relax
\mciteBstWouldAddEndPuncttrue
\mciteSetBstMidEndSepPunct{\mcitedefaultmidpunct}
{\mcitedefaultendpunct}{\mcitedefaultseppunct}\relax
\EndOfBibitem
\bibitem[{\citenamefont{Balian and Bloch}(1970)}]{BB1970}
\bibinfo{author}{\bibfnamefont{R.}~\bibnamefont{Balian}} \bibnamefont{and} \bibinfo{author}{\bibfnamefont{C.}~\bibnamefont{Bloch}}, \bibinfo{journal}{Annals of Physics} \textbf{\bibinfo{volume}{60}}, \bibinfo{pages}{401} (\bibinfo{year}{1970})\relax
\mciteBstWouldAddEndPuncttrue
\mciteSetBstMidEndSepPunct{\mcitedefaultmidpunct}
{\mcitedefaultendpunct}{\mcitedefaultseppunct}\relax
\EndOfBibitem
\bibitem[{\citenamefont{Balian and Bloch}(1972)}]{BB1972}
\bibinfo{author}{\bibfnamefont{R.}~\bibnamefont{Balian}} \bibnamefont{and} \bibinfo{author}{\bibfnamefont{C.}~\bibnamefont{Bloch}}, \textbf{\bibinfo{volume}{69}}, \bibinfo{pages}{76} (\bibinfo{year}{1972})\relax
\mciteBstWouldAddEndPuncttrue
\mciteSetBstMidEndSepPunct{\mcitedefaultmidpunct}
{\mcitedefaultendpunct}{\mcitedefaultseppunct}\relax
\EndOfBibitem
\bibitem[{\citenamefont{Balian and Bloch}(1974)}]{BB1974}
\bibinfo{author}{\bibfnamefont{R.}~\bibnamefont{Balian}} \bibnamefont{and} \bibinfo{author}{\bibfnamefont{C.}~\bibnamefont{Bloch}}, \bibinfo{journal}{Annals of Physics} \textbf{\bibinfo{volume}{85}}, \bibinfo{pages}{514} (\bibinfo{year}{1974})\relax
\mciteBstWouldAddEndPuncttrue
\mciteSetBstMidEndSepPunct{\mcitedefaultmidpunct}
{\mcitedefaultendpunct}{\mcitedefaultseppunct}\relax
\EndOfBibitem
\bibitem[{\citenamefont{Gutzwiller}(2013)}]{Gutb}
\bibinfo{author}{\bibfnamefont{M.~C.} \bibnamefont{Gutzwiller}}, \emph{\bibinfo{title}{Chaos in classical and quantum mechanics}}, vol.~\bibinfo{volume}{1} (\bibinfo{publisher}{Springer Science \& Business Media}, \bibinfo{year}{2013})\relax
\mciteBstWouldAddEndPuncttrue
\mciteSetBstMidEndSepPunct{\mcitedefaultmidpunct}
{\mcitedefaultendpunct}{\mcitedefaultseppunct}\relax
\EndOfBibitem
\bibitem[{\citenamefont{Schoeppl et~al.}(2024)\citenamefont{Schoeppl, Dubertrand, Urbina, and Richter}}]{suppmat}
\bibinfo{author}{\bibfnamefont{F.}~\bibnamefont{Schoeppl}}, \bibinfo{author}{\bibfnamefont{R.}~\bibnamefont{Dubertrand}}, \bibinfo{author}{\bibfnamefont{J.~D.} \bibnamefont{Urbina}}, \bibnamefont{and} \bibinfo{author}{\bibfnamefont{K.}~\bibnamefont{Richter}}, \bibinfo{journal}{Supplemental material}  (\bibinfo{year}{2024})\relax
\mciteBstWouldAddEndPuncttrue
\mciteSetBstMidEndSepPunct{\mcitedefaultmidpunct}
{\mcitedefaultendpunct}{\mcitedefaultseppunct}\relax
\EndOfBibitem
\bibitem[{Note1()}]{Note1}
Note1, \bibinfo{note}{following standard terminology we use ''single particle'' instead of the more precise ''first-quantized'' nomenclature for systems where the classical limit corresponds to particle-like (instead of field-like) degrees of freedom}\relax
\mciteBstWouldAddEndPuncttrue
\mciteSetBstMidEndSepPunct{\mcitedefaultmidpunct}
{\mcitedefaultendpunct}{\mcitedefaultseppunct}\relax
\EndOfBibitem
\bibitem[{\citenamefont{Bohigas et~al.}(1984)\citenamefont{Bohigas, Giannoni, and Schmit}}]{BGS}
\bibinfo{author}{\bibfnamefont{O.}~\bibnamefont{Bohigas}}, \bibinfo{author}{\bibfnamefont{M.~J.} \bibnamefont{Giannoni}}, \bibnamefont{and} \bibinfo{author}{\bibfnamefont{C.}~\bibnamefont{Schmit}}, \bibinfo{journal}{Physical Review Letters} \textbf{\bibinfo{volume}{52}}, \bibinfo{pages}{14} (\bibinfo{year}{1984})\relax
\mciteBstWouldAddEndPuncttrue
\mciteSetBstMidEndSepPunct{\mcitedefaultmidpunct}
{\mcitedefaultendpunct}{\mcitedefaultseppunct}\relax
\EndOfBibitem
\bibitem[{\citenamefont{Berry}(1985)}]{Berr3}
\bibinfo{author}{\bibfnamefont{M.~V.} \bibnamefont{Berry}}, \bibinfo{journal}{Proceedings of the Royal Society of London. A. Mathematical and Physical Sciences} \textbf{\bibinfo{volume}{400}}, \bibinfo{pages}{229} (\bibinfo{year}{1985})\relax
\mciteBstWouldAddEndPuncttrue
\mciteSetBstMidEndSepPunct{\mcitedefaultmidpunct}
{\mcitedefaultendpunct}{\mcitedefaultseppunct}\relax
\EndOfBibitem
\bibitem[{\citenamefont{Sieber and Richter}(2001)}]{Klaus1}
\bibinfo{author}{\bibfnamefont{M.}~\bibnamefont{Sieber}} \bibnamefont{and} \bibinfo{author}{\bibfnamefont{K.}~\bibnamefont{Richter}}, \bibinfo{journal}{Physica Scripta} \textbf{\bibinfo{volume}{T90}}, \bibinfo{pages}{128} (\bibinfo{year}{2001})\relax
\mciteBstWouldAddEndPuncttrue
\mciteSetBstMidEndSepPunct{\mcitedefaultmidpunct}
{\mcitedefaultendpunct}{\mcitedefaultseppunct}\relax
\EndOfBibitem
\bibitem[{\citenamefont{M\"uller et~al.}(2004)\citenamefont{M\"uller, Heusler, Braun, Haake, and Altland}}]{Haak1}
\bibinfo{author}{\bibfnamefont{S.}~\bibnamefont{M\"uller}}, \bibinfo{author}{\bibfnamefont{S.}~\bibnamefont{Heusler}}, \bibinfo{author}{\bibfnamefont{P.}~\bibnamefont{Braun}}, \bibinfo{author}{\bibfnamefont{F.}~\bibnamefont{Haake}}, \bibnamefont{and} \bibinfo{author}{\bibfnamefont{A.}~\bibnamefont{Altland}}, \bibinfo{journal}{Physical Review Letters} \textbf{\bibinfo{volume}{93}}, \bibinfo{pages}{014103} (\bibinfo{year}{2004})\relax
\mciteBstWouldAddEndPuncttrue
\mciteSetBstMidEndSepPunct{\mcitedefaultmidpunct}
{\mcitedefaultendpunct}{\mcitedefaultseppunct}\relax
\EndOfBibitem
\bibitem[{\citenamefont{M\"uller et~al.}(2005)\citenamefont{M\"uller, Heusler, Braun, Haake, and Altland}}]{Haak2}
\bibinfo{author}{\bibfnamefont{S.}~\bibnamefont{M\"uller}}, \bibinfo{author}{\bibfnamefont{S.}~\bibnamefont{Heusler}}, \bibinfo{author}{\bibfnamefont{P.}~\bibnamefont{Braun}}, \bibinfo{author}{\bibfnamefont{F.}~\bibnamefont{Haake}}, \bibnamefont{and} \bibinfo{author}{\bibfnamefont{A.}~\bibnamefont{Altland}}, \bibinfo{journal}{Physical Review E} \textbf{\bibinfo{volume}{72}}, \bibinfo{pages}{046207} (\bibinfo{year}{2005})\relax
\mciteBstWouldAddEndPuncttrue
\mciteSetBstMidEndSepPunct{\mcitedefaultmidpunct}
{\mcitedefaultendpunct}{\mcitedefaultseppunct}\relax
\EndOfBibitem
\bibitem[{\citenamefont{Brody et~al.}(1981)\citenamefont{Brody, Flores, French, Mello, Pandey, and Wong}}]{RevModPhys.53.385}
\bibinfo{author}{\bibfnamefont{T.~A.} \bibnamefont{Brody}}, \bibinfo{author}{\bibfnamefont{J.}~\bibnamefont{Flores}}, \bibinfo{author}{\bibfnamefont{J.~B.} \bibnamefont{French}}, \bibinfo{author}{\bibfnamefont{P.~A.} \bibnamefont{Mello}}, \bibinfo{author}{\bibfnamefont{A.}~\bibnamefont{Pandey}}, \bibnamefont{and} \bibinfo{author}{\bibfnamefont{S.~S.~M.} \bibnamefont{Wong}}, \bibinfo{journal}{Review of Modern Physics} \textbf{\bibinfo{volume}{53}}, \bibinfo{pages}{385} (\bibinfo{year}{1981})\relax
\mciteBstWouldAddEndPuncttrue
\mciteSetBstMidEndSepPunct{\mcitedefaultmidpunct}
{\mcitedefaultendpunct}{\mcitedefaultseppunct}\relax
\EndOfBibitem
\bibitem[{\citenamefont{Montambaux et~al.}(1993)\citenamefont{Montambaux, Poilblanc, Bellissard, and Sire}}]{montambaux_quantum_1993}
\bibinfo{author}{\bibfnamefont{G.}~\bibnamefont{Montambaux}}, \bibinfo{author}{\bibfnamefont{D.}~\bibnamefont{Poilblanc}}, \bibinfo{author}{\bibfnamefont{J.}~\bibnamefont{Bellissard}}, \bibnamefont{and} \bibinfo{author}{\bibfnamefont{C.}~\bibnamefont{Sire}}, \bibinfo{journal}{Physical Review Letters} \textbf{\bibinfo{volume}{70}}, \bibinfo{pages}{497} (\bibinfo{year}{1993})\relax
\mciteBstWouldAddEndPuncttrue
\mciteSetBstMidEndSepPunct{\mcitedefaultmidpunct}
{\mcitedefaultendpunct}{\mcitedefaultseppunct}\relax
\EndOfBibitem
\bibitem[{\citenamefont{Beenakker}(1997)}]{RevModPhys.69.731}
\bibinfo{author}{\bibfnamefont{C.~W.~J.} \bibnamefont{Beenakker}}, \bibinfo{journal}{Review of Modern Physics} \textbf{\bibinfo{volume}{69}}, \bibinfo{pages}{731} (\bibinfo{year}{1997})\relax
\mciteBstWouldAddEndPuncttrue
\mciteSetBstMidEndSepPunct{\mcitedefaultmidpunct}
{\mcitedefaultendpunct}{\mcitedefaultseppunct}\relax
\EndOfBibitem
\bibitem[{\citenamefont{Guhr et~al.}(1998)\citenamefont{Guhr, Müller–Groeling, and Weidenmüller}}]{GUHR1998189}
\bibinfo{author}{\bibfnamefont{T.}~\bibnamefont{Guhr}}, \bibinfo{author}{\bibfnamefont{A.}~\bibnamefont{Müller–Groeling}}, \bibnamefont{and} \bibinfo{author}{\bibfnamefont{H.~A.} \bibnamefont{Weidenmüller}}, \bibinfo{journal}{Physics Reports} \textbf{\bibinfo{volume}{299}}, \bibinfo{pages}{189} (\bibinfo{year}{1998})\relax
\mciteBstWouldAddEndPuncttrue
\mciteSetBstMidEndSepPunct{\mcitedefaultmidpunct}
{\mcitedefaultendpunct}{\mcitedefaultseppunct}\relax
\EndOfBibitem
\bibitem[{\citenamefont{Kolovsky and Buchleitner}(2004)}]{MBQC1}
\bibinfo{author}{\bibfnamefont{A.~R.} \bibnamefont{Kolovsky}} \bibnamefont{and} \bibinfo{author}{\bibfnamefont{A.}~\bibnamefont{Buchleitner}}, \bibinfo{journal}{Europhysics Letters} \textbf{\bibinfo{volume}{68}}, \bibinfo{pages}{632} (\bibinfo{year}{2004})\relax
\mciteBstWouldAddEndPuncttrue
\mciteSetBstMidEndSepPunct{\mcitedefaultmidpunct}
{\mcitedefaultendpunct}{\mcitedefaultseppunct}\relax
\EndOfBibitem
\bibitem[{\citenamefont{Kollath et~al.}(2010)\citenamefont{Kollath, Roux, Biroli, and Läuchli}}]{Kollath_2010}
\bibinfo{author}{\bibfnamefont{C.}~\bibnamefont{Kollath}}, \bibinfo{author}{\bibfnamefont{G.}~\bibnamefont{Roux}}, \bibinfo{author}{\bibfnamefont{G.}~\bibnamefont{Biroli}}, \bibnamefont{and} \bibinfo{author}{\bibfnamefont{A.~M.} \bibnamefont{Läuchli}}, \bibinfo{journal}{Journal of Statistical Mechanics: Theory and Experiment} \textbf{\bibinfo{volume}{2010}}, \bibinfo{pages}{P08011} (\bibinfo{year}{2010})\relax
\mciteBstWouldAddEndPuncttrue
\mciteSetBstMidEndSepPunct{\mcitedefaultmidpunct}
{\mcitedefaultendpunct}{\mcitedefaultseppunct}\relax
\EndOfBibitem
\bibitem[{\citenamefont{Santos and Rigol}(2010)}]{PhysRevE.81.036206}
\bibinfo{author}{\bibfnamefont{L.~F.} \bibnamefont{Santos}} \bibnamefont{and} \bibinfo{author}{\bibfnamefont{M.}~\bibnamefont{Rigol}}, \bibinfo{journal}{Physical Review E} \textbf{\bibinfo{volume}{81}}, \bibinfo{pages}{036206} (\bibinfo{year}{2010})\relax
\mciteBstWouldAddEndPuncttrue
\mciteSetBstMidEndSepPunct{\mcitedefaultmidpunct}
{\mcitedefaultendpunct}{\mcitedefaultseppunct}\relax
\EndOfBibitem
\bibitem[{\citenamefont{Richter et~al.}(2022)\citenamefont{Richter, Urbina, and Tomsovic}}]{Richter22}
\bibinfo{author}{\bibfnamefont{K.}~\bibnamefont{Richter}}, \bibinfo{author}{\bibfnamefont{J.~D.} \bibnamefont{Urbina}}, \bibnamefont{and} \bibinfo{author}{\bibfnamefont{S.}~\bibnamefont{Tomsovic}}, \bibinfo{journal}{Journal of Physics A: Mathematical and Theoretical} \textbf{\bibinfo{volume}{55}}, \bibinfo{pages}{453001} (\bibinfo{year}{2022})\relax
\mciteBstWouldAddEndPuncttrue
\mciteSetBstMidEndSepPunct{\mcitedefaultmidpunct}
{\mcitedefaultendpunct}{\mcitedefaultseppunct}\relax
\EndOfBibitem
\bibitem[{\citenamefont{Prosen}(2000)}]{10.1143/PTPS.139.191}
\bibinfo{author}{\bibfnamefont{T.}~\bibnamefont{Prosen}}, \bibinfo{journal}{Progress of Theoretical Physics Supplement} \textbf{\bibinfo{volume}{139}}, \bibinfo{pages}{191} (\bibinfo{year}{2000})\relax
\mciteBstWouldAddEndPuncttrue
\mciteSetBstMidEndSepPunct{\mcitedefaultmidpunct}
{\mcitedefaultendpunct}{\mcitedefaultseppunct}\relax
\EndOfBibitem
\bibitem[{\citenamefont{Rigol et~al.}(2008)\citenamefont{Rigol, Dunjko, and Olshanii}}]{Rig1}
\bibinfo{author}{\bibfnamefont{M.}~\bibnamefont{Rigol}}, \bibinfo{author}{\bibfnamefont{V.}~\bibnamefont{Dunjko}}, \bibnamefont{and} \bibinfo{author}{\bibfnamefont{M.}~\bibnamefont{Olshanii}}, \bibinfo{journal}{Nature} \textbf{\bibinfo{volume}{452}}, \bibinfo{pages}{854} (\bibinfo{year}{2008})\relax
\mciteBstWouldAddEndPuncttrue
\mciteSetBstMidEndSepPunct{\mcitedefaultmidpunct}
{\mcitedefaultendpunct}{\mcitedefaultseppunct}\relax
\EndOfBibitem
\bibitem[{\citenamefont{\ifmmode~\check{S}\else \v{S}\fi{}untajs et~al.}(2020)\citenamefont{\ifmmode~\check{S}\else \v{S}\fi{}untajs, Bon\ifmmode~\check{c}\else \v{c}\fi{}a, Prosen, and Vidmar}}]{PhysRevE.102.062144}
\bibinfo{author}{\bibfnamefont{J.}~\bibnamefont{\ifmmode~\check{S}\else \v{S}\fi{}untajs}}, \bibinfo{author}{\bibfnamefont{J.}~\bibnamefont{Bon\ifmmode~\check{c}\else \v{c}\fi{}a}}, \bibinfo{author}{\bibfnamefont{T.}~\bibnamefont{Prosen}}, \bibnamefont{and} \bibinfo{author}{\bibfnamefont{L.}~\bibnamefont{Vidmar}}, \bibinfo{journal}{Physical Review E} \textbf{\bibinfo{volume}{102}}, \bibinfo{pages}{062144} (\bibinfo{year}{2020})\relax
\mciteBstWouldAddEndPuncttrue
\mciteSetBstMidEndSepPunct{\mcitedefaultmidpunct}
{\mcitedefaultendpunct}{\mcitedefaultseppunct}\relax
\EndOfBibitem
\bibitem[{\citenamefont{Bertini et~al.}(2018)\citenamefont{Bertini, Kos, and Prosen}}]{PhysRevLett.121.264101}
\bibinfo{author}{\bibfnamefont{B.}~\bibnamefont{Bertini}}, \bibinfo{author}{\bibfnamefont{P.}~\bibnamefont{Kos}}, \bibnamefont{and} \bibinfo{author}{\bibfnamefont{T.}~\bibnamefont{Prosen}}, \bibinfo{journal}{Physical Review Letters} \textbf{\bibinfo{volume}{121}}, \bibinfo{pages}{264101} (\bibinfo{year}{2018})\relax
\mciteBstWouldAddEndPuncttrue
\mciteSetBstMidEndSepPunct{\mcitedefaultmidpunct}
{\mcitedefaultendpunct}{\mcitedefaultseppunct}\relax
\EndOfBibitem
\bibitem[{\citenamefont{Luitz et~al.}(2015)\citenamefont{Luitz, Laflorencie, and Alet}}]{PhysRevB.91.081103}
\bibinfo{author}{\bibfnamefont{D.~J.} \bibnamefont{Luitz}}, \bibinfo{author}{\bibfnamefont{N.}~\bibnamefont{Laflorencie}}, \bibnamefont{and} \bibinfo{author}{\bibfnamefont{F.}~\bibnamefont{Alet}}, \bibinfo{journal}{Physical Review B} \textbf{\bibinfo{volume}{91}}, \bibinfo{pages}{081103} (\bibinfo{year}{2015})\relax
\mciteBstWouldAddEndPuncttrue
\mciteSetBstMidEndSepPunct{\mcitedefaultmidpunct}
{\mcitedefaultendpunct}{\mcitedefaultseppunct}\relax
\EndOfBibitem
\bibitem[{\citenamefont{Kos et~al.}(2018)\citenamefont{Kos, Ljubotina, and Prosen}}]{Kos2018}
\bibinfo{author}{\bibfnamefont{P.}~\bibnamefont{Kos}}, \bibinfo{author}{\bibfnamefont{M.}~\bibnamefont{Ljubotina}}, \bibnamefont{and} \bibinfo{author}{\bibfnamefont{T.}~\bibnamefont{Prosen}}, \bibinfo{journal}{Phys. Rev. X} \textbf{\bibinfo{volume}{8}}, \bibinfo{pages}{021062} (\bibinfo{year}{2018})\relax
\mciteBstWouldAddEndPuncttrue
\mciteSetBstMidEndSepPunct{\mcitedefaultmidpunct}
{\mcitedefaultendpunct}{\mcitedefaultseppunct}\relax
\EndOfBibitem
\bibitem[{\citenamefont{McDonald and Kaufman}(1979)}]{mcdonaldSpectrumEigenfunctionsHamiltonian1979}
\bibinfo{author}{\bibfnamefont{S.~W.} \bibnamefont{McDonald}} \bibnamefont{and} \bibinfo{author}{\bibfnamefont{A.~N.} \bibnamefont{Kaufman}}, \bibinfo{journal}{Physical Review Letters} \textbf{\bibinfo{volume}{42}}, \bibinfo{pages}{1189} (\bibinfo{year}{1979})\relax
\mciteBstWouldAddEndPuncttrue
\mciteSetBstMidEndSepPunct{\mcitedefaultmidpunct}
{\mcitedefaultendpunct}{\mcitedefaultseppunct}\relax
\EndOfBibitem
\bibitem[{\citenamefont{McDonald and Kaufman}(1988{\natexlab{a}})}]{mcdonaldWaveChaosStadium1988}
\bibinfo{author}{\bibfnamefont{S.~W.} \bibnamefont{McDonald}} \bibnamefont{and} \bibinfo{author}{\bibfnamefont{A.~N.} \bibnamefont{Kaufman}}, \bibinfo{journal}{Physical Review A} \textbf{\bibinfo{volume}{37}}, \bibinfo{pages}{3067} (\bibinfo{year}{1988}{\natexlab{a}})\relax
\mciteBstWouldAddEndPuncttrue
\mciteSetBstMidEndSepPunct{\mcitedefaultmidpunct}
{\mcitedefaultendpunct}{\mcitedefaultseppunct}\relax
\EndOfBibitem
\bibitem[{\citenamefont{Berry}(1977{\natexlab{a}})}]{Berr4}
\bibinfo{author}{\bibfnamefont{M.~V.} \bibnamefont{Berry}}, \bibinfo{journal}{Journal of Physics A: Mathematical and General} \textbf{\bibinfo{volume}{10}}, \bibinfo{pages}{2083} (\bibinfo{year}{1977}{\natexlab{a}})\relax
\mciteBstWouldAddEndPuncttrue
\mciteSetBstMidEndSepPunct{\mcitedefaultmidpunct}
{\mcitedefaultendpunct}{\mcitedefaultseppunct}\relax
\EndOfBibitem
\bibitem[{\citenamefont{Hortikar and Srednicki}(1998)}]{Sied1}
\bibinfo{author}{\bibfnamefont{S.}~\bibnamefont{Hortikar}} \bibnamefont{and} \bibinfo{author}{\bibfnamefont{M.}~\bibnamefont{Srednicki}}, \bibinfo{journal}{Physical Review Letters} \textbf{\bibinfo{volume}{80}}, \bibinfo{pages}{1646} (\bibinfo{year}{1998})\relax
\mciteBstWouldAddEndPuncttrue
\mciteSetBstMidEndSepPunct{\mcitedefaultmidpunct}
{\mcitedefaultendpunct}{\mcitedefaultseppunct}\relax
\EndOfBibitem
\bibitem[{\citenamefont{Urbina and Richter}(2013)}]{JDUre}
\bibinfo{author}{\bibfnamefont{J.~D.} \bibnamefont{Urbina}} \bibnamefont{and} \bibinfo{author}{\bibfnamefont{K.}~\bibnamefont{Richter}}, \bibinfo{journal}{Advances in Physics} \textbf{\bibinfo{volume}{62}}, \bibinfo{pages}{363} (\bibinfo{year}{2013})\relax
\mciteBstWouldAddEndPuncttrue
\mciteSetBstMidEndSepPunct{\mcitedefaultmidpunct}
{\mcitedefaultendpunct}{\mcitedefaultseppunct}\relax
\EndOfBibitem
\bibitem[{\citenamefont{Chang et~al.}(1996)\citenamefont{Chang, Baranger, Pfeiffer, West, and Chang}}]{RWM3}
\bibinfo{author}{\bibfnamefont{A.~M.} \bibnamefont{Chang}}, \bibinfo{author}{\bibfnamefont{H.~U.} \bibnamefont{Baranger}}, \bibinfo{author}{\bibfnamefont{L.~N.} \bibnamefont{Pfeiffer}}, \bibinfo{author}{\bibfnamefont{K.~W.} \bibnamefont{West}}, \bibnamefont{and} \bibinfo{author}{\bibfnamefont{T.~Y.} \bibnamefont{Chang}}, \bibinfo{journal}{Physical Review Letters} \textbf{\bibinfo{volume}{76}}, \bibinfo{pages}{1695} (\bibinfo{year}{1996})\relax
\mciteBstWouldAddEndPuncttrue
\mciteSetBstMidEndSepPunct{\mcitedefaultmidpunct}
{\mcitedefaultendpunct}{\mcitedefaultseppunct}\relax
\EndOfBibitem
\bibitem[{\citenamefont{Urbina and Richter}(2006)}]{Urbina06}
\bibinfo{author}{\bibfnamefont{J.~D.} \bibnamefont{Urbina}} \bibnamefont{and} \bibinfo{author}{\bibfnamefont{K.}~\bibnamefont{Richter}}, \bibinfo{journal}{Physical Review Letters} \textbf{\bibinfo{volume}{97}}, \bibinfo{pages}{214101} (\bibinfo{year}{2006})\relax
\mciteBstWouldAddEndPuncttrue
\mciteSetBstMidEndSepPunct{\mcitedefaultmidpunct}
{\mcitedefaultendpunct}{\mcitedefaultseppunct}\relax
\EndOfBibitem
\bibitem[{\citenamefont{Dennis}(2007)}]{RWM2}
\bibinfo{author}{\bibfnamefont{M.~R.} \bibnamefont{Dennis}}, \bibinfo{journal}{The European Physical Journal Special Topics} \textbf{\bibinfo{volume}{145}}, \bibinfo{pages}{191} (\bibinfo{year}{2007})\relax
\mciteBstWouldAddEndPuncttrue
\mciteSetBstMidEndSepPunct{\mcitedefaultmidpunct}
{\mcitedefaultendpunct}{\mcitedefaultseppunct}\relax
\EndOfBibitem
\bibitem[{\citenamefont{McDonald and Kaufman}(1988{\natexlab{b}})}]{PhysRevA.37.3067}
\bibinfo{author}{\bibfnamefont{S.~W.} \bibnamefont{McDonald}} \bibnamefont{and} \bibinfo{author}{\bibfnamefont{A.~N.} \bibnamefont{Kaufman}}, \bibinfo{journal}{Physical Review A} \textbf{\bibinfo{volume}{37}}, \bibinfo{pages}{3067} (\bibinfo{year}{1988}{\natexlab{b}})\relax
\mciteBstWouldAddEndPuncttrue
\mciteSetBstMidEndSepPunct{\mcitedefaultmidpunct}
{\mcitedefaultendpunct}{\mcitedefaultseppunct}\relax
\EndOfBibitem
\bibitem[{\citenamefont{Berry and Dennis}(2000)}]{RWMBe3}
\bibinfo{author}{\bibfnamefont{M.~V.} \bibnamefont{Berry}} \bibnamefont{and} \bibinfo{author}{\bibfnamefont{M.~R.} \bibnamefont{Dennis}}, \bibinfo{journal}{Proceedings of the Royal Society of London. Series A: Mathematical, Physical and Engineering Sciences} \textbf{\bibinfo{volume}{456}}, \bibinfo{pages}{2059} (\bibinfo{year}{2000})\relax
\mciteBstWouldAddEndPuncttrue
\mciteSetBstMidEndSepPunct{\mcitedefaultmidpunct}
{\mcitedefaultendpunct}{\mcitedefaultseppunct}\relax
\EndOfBibitem
\bibitem[{\citenamefont{Bogomolny and Schmit}(2002)}]{bogomolny2002percolation}
\bibinfo{author}{\bibfnamefont{E.}~\bibnamefont{Bogomolny}} \bibnamefont{and} \bibinfo{author}{\bibfnamefont{C.}~\bibnamefont{Schmit}}, \bibinfo{journal}{Physical Review Letters} \textbf{\bibinfo{volume}{88}}, \bibinfo{pages}{114102} (\bibinfo{year}{2002})\relax
\mciteBstWouldAddEndPuncttrue
\mciteSetBstMidEndSepPunct{\mcitedefaultmidpunct}
{\mcitedefaultendpunct}{\mcitedefaultseppunct}\relax
\EndOfBibitem
\bibitem[{\citenamefont{Barth and St\"ockmann}(2002)}]{HJ1}
\bibinfo{author}{\bibfnamefont{M.}~\bibnamefont{Barth}} \bibnamefont{and} \bibinfo{author}{\bibfnamefont{H.-J.} \bibnamefont{St\"ockmann}}, \bibinfo{journal}{Physical Review E} \textbf{\bibinfo{volume}{65}}, \bibinfo{pages}{066208} (\bibinfo{year}{2002})\relax
\mciteBstWouldAddEndPuncttrue
\mciteSetBstMidEndSepPunct{\mcitedefaultmidpunct}
{\mcitedefaultendpunct}{\mcitedefaultseppunct}\relax
\EndOfBibitem
\bibitem[{\citenamefont{B\"arnthaler et~al.}(2010)\citenamefont{B\"arnthaler, Rotter, Libisch, Burgd\"orfer, Gehler, Kuhl, and St\"ockmann}}]{HJ2}
\bibinfo{author}{\bibfnamefont{A.}~\bibnamefont{B\"arnthaler}}, \bibinfo{author}{\bibfnamefont{S.}~\bibnamefont{Rotter}}, \bibinfo{author}{\bibfnamefont{F.}~\bibnamefont{Libisch}}, \bibinfo{author}{\bibfnamefont{J.}~\bibnamefont{Burgd\"orfer}}, \bibinfo{author}{\bibfnamefont{S.}~\bibnamefont{Gehler}}, \bibinfo{author}{\bibfnamefont{U.}~\bibnamefont{Kuhl}}, \bibnamefont{and} \bibinfo{author}{\bibfnamefont{H.-J.} \bibnamefont{St\"ockmann}}, \bibinfo{journal}{Physical Review Letters} \textbf{\bibinfo{volume}{105}}, \bibinfo{pages}{056801} (\bibinfo{year}{2010})\relax
\mciteBstWouldAddEndPuncttrue
\mciteSetBstMidEndSepPunct{\mcitedefaultmidpunct}
{\mcitedefaultendpunct}{\mcitedefaultseppunct}\relax
\EndOfBibitem
\bibitem[{\citenamefont{St{\"o}ckmann}(2013)}]{HJ3}
\bibinfo{author}{\bibfnamefont{H.-J.} \bibnamefont{St{\"o}ckmann}}, \emph{\bibinfo{title}{Chaos in Microwave resonators}} (\bibinfo{publisher}{Springer}, \bibinfo{year}{2013})\relax
\mciteBstWouldAddEndPuncttrue
\mciteSetBstMidEndSepPunct{\mcitedefaultmidpunct}
{\mcitedefaultendpunct}{\mcitedefaultseppunct}\relax
\EndOfBibitem
\bibitem[{\citenamefont{Berry and Dennis}(2001)}]{RWMBe4}
\bibinfo{author}{\bibfnamefont{M.}~\bibnamefont{Berry}} \bibnamefont{and} \bibinfo{author}{\bibfnamefont{M.}~\bibnamefont{Dennis}}, \bibinfo{journal}{Proceedings of the Royal Society of London. Series A: Mathematical, Physical and Engineering Sciences} \textbf{\bibinfo{volume}{457}}, \bibinfo{pages}{141} (\bibinfo{year}{2001})\relax
\mciteBstWouldAddEndPuncttrue
\mciteSetBstMidEndSepPunct{\mcitedefaultmidpunct}
{\mcitedefaultendpunct}{\mcitedefaultseppunct}\relax
\EndOfBibitem
\bibitem[{\citenamefont{H{\"o}hmann et~al.}(2009)\citenamefont{H{\"o}hmann, Kuhl, St{\"o}ckmann, Urbina, and Dennis}}]{Dennis1}
\bibinfo{author}{\bibfnamefont{R.}~\bibnamefont{H{\"o}hmann}}, \bibinfo{author}{\bibfnamefont{U.}~\bibnamefont{Kuhl}}, \bibinfo{author}{\bibfnamefont{H.-J.} \bibnamefont{St{\"o}ckmann}}, \bibinfo{author}{\bibfnamefont{J.}~\bibnamefont{Urbina}}, \bibnamefont{and} \bibinfo{author}{\bibfnamefont{M.}~\bibnamefont{Dennis}}, \bibinfo{journal}{Physical Review E} \textbf{\bibinfo{volume}{79}}, \bibinfo{pages}{016203} (\bibinfo{year}{2009})\relax
\mciteBstWouldAddEndPuncttrue
\mciteSetBstMidEndSepPunct{\mcitedefaultmidpunct}
{\mcitedefaultendpunct}{\mcitedefaultseppunct}\relax
\EndOfBibitem
\bibitem[{\citenamefont{Taylor and Dennis}(2016)}]{Dennis2}
\bibinfo{author}{\bibfnamefont{A.~J.} \bibnamefont{Taylor}} \bibnamefont{and} \bibinfo{author}{\bibfnamefont{M.~R.} \bibnamefont{Dennis}}, \bibinfo{journal}{Nature communications} \textbf{\bibinfo{volume}{7}}, \bibinfo{pages}{1} (\bibinfo{year}{2016})\relax
\mciteBstWouldAddEndPuncttrue
\mciteSetBstMidEndSepPunct{\mcitedefaultmidpunct}
{\mcitedefaultendpunct}{\mcitedefaultseppunct}\relax
\EndOfBibitem
\bibitem[{\citenamefont{Jain and Samajdar}(2017)}]{jainNodalPortraitsQuantum2017}
\bibinfo{author}{\bibfnamefont{S.~R.} \bibnamefont{Jain}} \bibnamefont{and} \bibinfo{author}{\bibfnamefont{R.}~\bibnamefont{Samajdar}}, \bibinfo{journal}{Reviews of Modern Physics} \textbf{\bibinfo{volume}{89}} (\bibinfo{year}{2017})\relax
\mciteBstWouldAddEndPuncttrue
\mciteSetBstMidEndSepPunct{\mcitedefaultmidpunct}
{\mcitedefaultendpunct}{\mcitedefaultseppunct}\relax
\EndOfBibitem
\bibitem[{\citenamefont{Lubkin}(1978)}]{lubkinEntropyNsystemIts1978}
\bibinfo{author}{\bibfnamefont{E.}~\bibnamefont{Lubkin}}, \bibinfo{journal}{Journal of Mathematical Physics} \textbf{\bibinfo{volume}{19}}, \bibinfo{pages}{1028} (\bibinfo{year}{1978})\relax
\mciteBstWouldAddEndPuncttrue
\mciteSetBstMidEndSepPunct{\mcitedefaultmidpunct}
{\mcitedefaultendpunct}{\mcitedefaultseppunct}\relax
\EndOfBibitem
\bibitem[{\citenamefont{Page}(1993)}]{pageAverageEntropySubsystem1993}
\bibinfo{author}{\bibfnamefont{D.~N.} \bibnamefont{Page}}, \bibinfo{journal}{Physical Review Letters} \textbf{\bibinfo{volume}{71}}, \bibinfo{pages}{1291} (\bibinfo{year}{1993})\relax
\mciteBstWouldAddEndPuncttrue
\mciteSetBstMidEndSepPunct{\mcitedefaultmidpunct}
{\mcitedefaultendpunct}{\mcitedefaultseppunct}\relax
\EndOfBibitem
\bibitem[{\citenamefont{Sen}(1996)}]{senAverageEntropyQuantum1996}
\bibinfo{author}{\bibfnamefont{S.}~\bibnamefont{Sen}}, \bibinfo{journal}{Physical Review Letters} \textbf{\bibinfo{volume}{77}}, \bibinfo{pages}{1} (\bibinfo{year}{1996})\relax
\mciteBstWouldAddEndPuncttrue
\mciteSetBstMidEndSepPunct{\mcitedefaultmidpunct}
{\mcitedefaultendpunct}{\mcitedefaultseppunct}\relax
\EndOfBibitem
\bibitem[{\citenamefont{Deutsch}(1991)}]{ETH1}
\bibinfo{author}{\bibfnamefont{J.~M.} \bibnamefont{Deutsch}}, \bibinfo{journal}{Physical Review A} \textbf{\bibinfo{volume}{43}}, \bibinfo{pages}{2046} (\bibinfo{year}{1991})\relax
\mciteBstWouldAddEndPuncttrue
\mciteSetBstMidEndSepPunct{\mcitedefaultmidpunct}
{\mcitedefaultendpunct}{\mcitedefaultseppunct}\relax
\EndOfBibitem
\bibitem[{\citenamefont{Srednicki}(1994)}]{ETH2}
\bibinfo{author}{\bibfnamefont{M.}~\bibnamefont{Srednicki}}, \bibinfo{journal}{Physical Review E} \textbf{\bibinfo{volume}{50}}, \bibinfo{pages}{888} (\bibinfo{year}{1994})\relax
\mciteBstWouldAddEndPuncttrue
\mciteSetBstMidEndSepPunct{\mcitedefaultmidpunct}
{\mcitedefaultendpunct}{\mcitedefaultseppunct}\relax
\EndOfBibitem
\bibitem[{\citenamefont{D'Alessio et~al.}(2016)\citenamefont{D'Alessio, Kafri, Polkovnikov, and Rigol}}]{ETH3}
\bibinfo{author}{\bibfnamefont{L.}~\bibnamefont{D'Alessio}}, \bibinfo{author}{\bibfnamefont{Y.}~\bibnamefont{Kafri}}, \bibinfo{author}{\bibfnamefont{A.}~\bibnamefont{Polkovnikov}}, \bibnamefont{and} \bibinfo{author}{\bibfnamefont{M.}~\bibnamefont{Rigol}}, \bibinfo{journal}{Advances in Physics} \textbf{\bibinfo{volume}{65}}, \bibinfo{pages}{239} (\bibinfo{year}{2016})\relax
\mciteBstWouldAddEndPuncttrue
\mciteSetBstMidEndSepPunct{\mcitedefaultmidpunct}
{\mcitedefaultendpunct}{\mcitedefaultseppunct}\relax
\EndOfBibitem
\bibitem[{\citenamefont{Deutsch}(2018)}]{ETH4}
\bibinfo{author}{\bibfnamefont{J.~M.} \bibnamefont{Deutsch}}, \bibinfo{journal}{Reports on Progress in Physics} \textbf{\bibinfo{volume}{81}}, \bibinfo{pages}{082001} (\bibinfo{year}{2018})\relax
\mciteBstWouldAddEndPuncttrue
\mciteSetBstMidEndSepPunct{\mcitedefaultmidpunct}
{\mcitedefaultendpunct}{\mcitedefaultseppunct}\relax
\EndOfBibitem
\bibitem[{\citenamefont{Jain and Alonso}(1997)}]{jainQuantumChaosRandom1997}
\bibinfo{author}{\bibfnamefont{S.~R.} \bibnamefont{Jain}} \bibnamefont{and} \bibinfo{author}{\bibfnamefont{D.}~\bibnamefont{Alonso}}, \bibinfo{journal}{Journal of Physics A: Mathematical and General} \textbf{\bibinfo{volume}{30}}, \bibinfo{pages}{4993} (\bibinfo{year}{1997})\relax
\mciteBstWouldAddEndPuncttrue
\mciteSetBstMidEndSepPunct{\mcitedefaultmidpunct}
{\mcitedefaultendpunct}{\mcitedefaultseppunct}\relax
\EndOfBibitem
\bibitem[{\citenamefont{Alonso and Jain}(1996)}]{ALONSO1996812}
\bibinfo{author}{\bibfnamefont{D.}~\bibnamefont{Alonso}} \bibnamefont{and} \bibinfo{author}{\bibfnamefont{S.~R.} \bibnamefont{Jain}}, \bibinfo{journal}{Physics Letters B} \textbf{\bibinfo{volume}{387}}, \bibinfo{pages}{812} (\bibinfo{year}{1996})\relax
\mciteBstWouldAddEndPuncttrue
\mciteSetBstMidEndSepPunct{\mcitedefaultmidpunct}
{\mcitedefaultendpunct}{\mcitedefaultseppunct}\relax
\EndOfBibitem
\bibitem[{\citenamefont{Gaspard and Jain}(1997)}]{Gaspard1997}
\bibinfo{author}{\bibfnamefont{P.}~\bibnamefont{Gaspard}} \bibnamefont{and} \bibinfo{author}{\bibfnamefont{S.~R.} \bibnamefont{Jain}}, \bibinfo{journal}{Pramana} \textbf{\bibinfo{volume}{48}}, \bibinfo{pages}{503} (\bibinfo{year}{1997})\relax
\mciteBstWouldAddEndPuncttrue
\mciteSetBstMidEndSepPunct{\mcitedefaultmidpunct}
{\mcitedefaultendpunct}{\mcitedefaultseppunct}\relax
\EndOfBibitem
\bibitem[{\citenamefont{Gross and Bloch}(2017)}]{doi:10.1126/science.aal3837}
\bibinfo{author}{\bibfnamefont{C.}~\bibnamefont{Gross}} \bibnamefont{and} \bibinfo{author}{\bibfnamefont{I.}~\bibnamefont{Bloch}}, \bibinfo{journal}{Science} \textbf{\bibinfo{volume}{357}}, \bibinfo{pages}{995} (\bibinfo{year}{2017})\relax
\mciteBstWouldAddEndPuncttrue
\mciteSetBstMidEndSepPunct{\mcitedefaultmidpunct}
{\mcitedefaultendpunct}{\mcitedefaultseppunct}\relax
\EndOfBibitem
\bibitem[{\citenamefont{Aidelsburger}(2018)}]{Aidelsburger_2018}
\bibinfo{author}{\bibfnamefont{M.}~\bibnamefont{Aidelsburger}}, \bibinfo{journal}{Journal of Physics B: Atomic, Molecular and Optical Physics} \textbf{\bibinfo{volume}{51}}, \bibinfo{pages}{193001} (\bibinfo{year}{2018})\relax
\mciteBstWouldAddEndPuncttrue
\mciteSetBstMidEndSepPunct{\mcitedefaultmidpunct}
{\mcitedefaultendpunct}{\mcitedefaultseppunct}\relax
\EndOfBibitem
\bibitem[{\citenamefont{Bloch et~al.}(2008{\natexlab{a}})\citenamefont{Bloch, Dalibard, and Zwerger}}]{RevModPhys.80.885}
\bibinfo{author}{\bibfnamefont{I.}~\bibnamefont{Bloch}}, \bibinfo{author}{\bibfnamefont{J.}~\bibnamefont{Dalibard}}, \bibnamefont{and} \bibinfo{author}{\bibfnamefont{W.}~\bibnamefont{Zwerger}}, \bibinfo{journal}{Review of Modern Physics} \textbf{\bibinfo{volume}{80}}, \bibinfo{pages}{885} (\bibinfo{year}{2008}{\natexlab{a}})\relax
\mciteBstWouldAddEndPuncttrue
\mciteSetBstMidEndSepPunct{\mcitedefaultmidpunct}
{\mcitedefaultendpunct}{\mcitedefaultseppunct}\relax
\EndOfBibitem
\bibitem[{\citenamefont{Lanyon et~al.}(2017)\citenamefont{Lanyon, Maier, Holz{\"a}pfel, Baumgratz, Hempel, Jurcevic, Dhand, Buyskikh, Daley, Cramer et~al.}}]{Lanyon2017}
\bibinfo{author}{\bibfnamefont{B.~P.} \bibnamefont{Lanyon}}, \bibinfo{author}{\bibfnamefont{C.}~\bibnamefont{Maier}}, \bibinfo{author}{\bibfnamefont{M.}~\bibnamefont{Holz{\"a}pfel}}, \bibinfo{author}{\bibfnamefont{T.}~\bibnamefont{Baumgratz}}, \bibinfo{author}{\bibfnamefont{C.}~\bibnamefont{Hempel}}, \bibinfo{author}{\bibfnamefont{P.}~\bibnamefont{Jurcevic}}, \bibinfo{author}{\bibfnamefont{I.}~\bibnamefont{Dhand}}, \bibinfo{author}{\bibfnamefont{A.~S.} \bibnamefont{Buyskikh}}, \bibinfo{author}{\bibfnamefont{A.~J.} \bibnamefont{Daley}}, \bibinfo{author}{\bibfnamefont{M.}~\bibnamefont{Cramer}}, \bibnamefont{et~al.}, \bibinfo{journal}{Nature Physics} \textbf{\bibinfo{volume}{13}}, \bibinfo{pages}{1158} (\bibinfo{year}{2017})\relax
\mciteBstWouldAddEndPuncttrue
\mciteSetBstMidEndSepPunct{\mcitedefaultmidpunct}
{\mcitedefaultendpunct}{\mcitedefaultseppunct}\relax
\EndOfBibitem
\bibitem[{\citenamefont{Haake}(2006)}]{Haake06}
\bibinfo{author}{\bibfnamefont{F.}~\bibnamefont{Haake}}, \emph{\bibinfo{title}{Quantum Signatures of Chaos}} (\bibinfo{publisher}{Springer-Verlag}, \bibinfo{address}{Berlin, Heidelberg}, \bibinfo{year}{2006}), ISBN \bibinfo{isbn}{3540677232}\relax
\mciteBstWouldAddEndPuncttrue
\mciteSetBstMidEndSepPunct{\mcitedefaultmidpunct}
{\mcitedefaultendpunct}{\mcitedefaultseppunct}\relax
\EndOfBibitem
\bibitem[{\citenamefont{Ullmo}(2008)}]{MBQC3}
\bibinfo{author}{\bibfnamefont{D.}~\bibnamefont{Ullmo}}, \bibinfo{journal}{Reports on Progress in Physics} \textbf{\bibinfo{volume}{71}}, \bibinfo{pages}{026001} (\bibinfo{year}{2008})\relax
\mciteBstWouldAddEndPuncttrue
\mciteSetBstMidEndSepPunct{\mcitedefaultmidpunct}
{\mcitedefaultendpunct}{\mcitedefaultseppunct}\relax
\EndOfBibitem
\bibitem[{\citenamefont{Cotler et~al.}(2016)\citenamefont{Cotler, Gur-Ari, Hanada, Polchinski, Saad, Shenker, Stanford, Streicher, and Tezuka}}]{JT1}
\bibinfo{author}{\bibfnamefont{J.~S.} \bibnamefont{Cotler}}, \bibinfo{author}{\bibfnamefont{G.}~\bibnamefont{Gur-Ari}}, \bibinfo{author}{\bibfnamefont{M.}~\bibnamefont{Hanada}}, \bibinfo{author}{\bibfnamefont{J.}~\bibnamefont{Polchinski}}, \bibinfo{author}{\bibfnamefont{P.}~\bibnamefont{Saad}}, \bibinfo{author}{\bibfnamefont{S.~H.} \bibnamefont{Shenker}}, \bibinfo{author}{\bibfnamefont{D.}~\bibnamefont{Stanford}}, \bibinfo{author}{\bibfnamefont{A.}~\bibnamefont{Streicher}}, \bibnamefont{and} \bibinfo{author}{\bibfnamefont{M.}~\bibnamefont{Tezuka}}, \bibinfo{journal}{Journal of High Energy Physics} \textbf{\bibinfo{volume}{2017}}, \bibinfo{pages}{1} (\bibinfo{year}{2016})\relax
\mciteBstWouldAddEndPuncttrue
\mciteSetBstMidEndSepPunct{\mcitedefaultmidpunct}
{\mcitedefaultendpunct}{\mcitedefaultseppunct}\relax
\EndOfBibitem
\bibitem[{\citenamefont{Sadd et~al.}(2016)\citenamefont{Sadd, Shenker, and Stanford}}]{JT2}
\bibinfo{author}{\bibfnamefont{P.}~\bibnamefont{Sadd}}, \bibinfo{author}{\bibfnamefont{S.~H.} \bibnamefont{Shenker}}, \bibnamefont{and} \bibinfo{author}{\bibfnamefont{D.}~\bibnamefont{Stanford}}, \bibinfo{journal}{Journal of High Energy Physics} \textbf{\bibinfo{volume}{2017}}, \bibinfo{pages}{1} (\bibinfo{year}{2016})\relax
\mciteBstWouldAddEndPuncttrue
\mciteSetBstMidEndSepPunct{\mcitedefaultmidpunct}
{\mcitedefaultendpunct}{\mcitedefaultseppunct}\relax
\EndOfBibitem
\bibitem[{\citenamefont{Engl et~al.}(2014)\citenamefont{Engl, Dujardin, Arg\"uelles, Schlagheck, Richter, and Urbina}}]{CBS}
\bibinfo{author}{\bibfnamefont{T.}~\bibnamefont{Engl}}, \bibinfo{author}{\bibfnamefont{J.}~\bibnamefont{Dujardin}}, \bibinfo{author}{\bibfnamefont{A.}~\bibnamefont{Arg\"uelles}}, \bibinfo{author}{\bibfnamefont{P.}~\bibnamefont{Schlagheck}}, \bibinfo{author}{\bibfnamefont{K.}~\bibnamefont{Richter}}, \bibnamefont{and} \bibinfo{author}{\bibfnamefont{J.~D.} \bibnamefont{Urbina}}, \bibinfo{journal}{Physical Review Letters} \textbf{\bibinfo{volume}{112}}, \bibinfo{pages}{140403} (\bibinfo{year}{2014})\relax
\mciteBstWouldAddEndPuncttrue
\mciteSetBstMidEndSepPunct{\mcitedefaultmidpunct}
{\mcitedefaultendpunct}{\mcitedefaultseppunct}\relax
\EndOfBibitem
\bibitem[{\citenamefont{Engl et~al.}(2018)\citenamefont{Engl, Urbina, Richter, and Schlagheck}}]{MBSE}
\bibinfo{author}{\bibfnamefont{T.}~\bibnamefont{Engl}}, \bibinfo{author}{\bibfnamefont{J.~D.} \bibnamefont{Urbina}}, \bibinfo{author}{\bibfnamefont{K.}~\bibnamefont{Richter}}, \bibnamefont{and} \bibinfo{author}{\bibfnamefont{P.}~\bibnamefont{Schlagheck}}, \bibinfo{journal}{Physical Review A} \textbf{\bibinfo{volume}{98}}, \bibinfo{pages}{013630} (\bibinfo{year}{2018})\relax
\mciteBstWouldAddEndPuncttrue
\mciteSetBstMidEndSepPunct{\mcitedefaultmidpunct}
{\mcitedefaultendpunct}{\mcitedefaultseppunct}\relax
\EndOfBibitem
\bibitem[{\citenamefont{Engl et~al.}(2015)\citenamefont{Engl, Urbina, and Richter}}]{TF_Thomas}
\bibinfo{author}{\bibfnamefont{T.}~\bibnamefont{Engl}}, \bibinfo{author}{\bibfnamefont{J.~D.} \bibnamefont{Urbina}}, \bibnamefont{and} \bibinfo{author}{\bibfnamefont{K.}~\bibnamefont{Richter}}, \bibinfo{journal}{Physical Review E} \textbf{\bibinfo{volume}{92}}, \bibinfo{pages}{062907} (\bibinfo{year}{2015})\relax
\mciteBstWouldAddEndPuncttrue
\mciteSetBstMidEndSepPunct{\mcitedefaultmidpunct}
{\mcitedefaultendpunct}{\mcitedefaultseppunct}\relax
\EndOfBibitem
\bibitem[{\citenamefont{Dubertrand and Müller}(2016)}]{TF_Remy}
\bibinfo{author}{\bibfnamefont{R.}~\bibnamefont{Dubertrand}} \bibnamefont{and} \bibinfo{author}{\bibfnamefont{S.}~\bibnamefont{Müller}}, \bibinfo{journal}{New Journal of Physics} \textbf{\bibinfo{volume}{18}}, \bibinfo{pages}{033009} (\bibinfo{year}{2016})\relax
\mciteBstWouldAddEndPuncttrue
\mciteSetBstMidEndSepPunct{\mcitedefaultmidpunct}
{\mcitedefaultendpunct}{\mcitedefaultseppunct}\relax
\EndOfBibitem
\bibitem[{\citenamefont{Rammensee et~al.}(2018)\citenamefont{Rammensee, Urbina, and Richter}}]{Josef}
\bibinfo{author}{\bibfnamefont{J.}~\bibnamefont{Rammensee}}, \bibinfo{author}{\bibfnamefont{J.~D.} \bibnamefont{Urbina}}, \bibnamefont{and} \bibinfo{author}{\bibfnamefont{K.}~\bibnamefont{Richter}}, \bibinfo{journal}{Physical Review Letters} \textbf{\bibinfo{volume}{121}}, \bibinfo{pages}{124101} (\bibinfo{year}{2018})\relax
\mciteBstWouldAddEndPuncttrue
\mciteSetBstMidEndSepPunct{\mcitedefaultmidpunct}
{\mcitedefaultendpunct}{\mcitedefaultseppunct}\relax
\EndOfBibitem
\bibitem[{\citenamefont{Hummel et~al.}(2019)\citenamefont{Hummel, Geiger, Urbina, and Richter}}]{QB}
\bibinfo{author}{\bibfnamefont{Q.}~\bibnamefont{Hummel}}, \bibinfo{author}{\bibfnamefont{B.}~\bibnamefont{Geiger}}, \bibinfo{author}{\bibfnamefont{J.~D.} \bibnamefont{Urbina}}, \bibnamefont{and} \bibinfo{author}{\bibfnamefont{K.}~\bibnamefont{Richter}}, \bibinfo{journal}{Physical Review Letters} \textbf{\bibinfo{volume}{123}}, \bibinfo{pages}{160401} (\bibinfo{year}{2019})\relax
\mciteBstWouldAddEndPuncttrue
\mciteSetBstMidEndSepPunct{\mcitedefaultmidpunct}
{\mcitedefaultendpunct}{\mcitedefaultseppunct}\relax
\EndOfBibitem
\bibitem[{\citenamefont{Tomsovic et~al.}(2018)\citenamefont{Tomsovic, Schlagheck, Ullmo, Urbina, and Richter}}]{PhysRevA.97.061606}
\bibinfo{author}{\bibfnamefont{S.}~\bibnamefont{Tomsovic}}, \bibinfo{author}{\bibfnamefont{P.}~\bibnamefont{Schlagheck}}, \bibinfo{author}{\bibfnamefont{D.}~\bibnamefont{Ullmo}}, \bibinfo{author}{\bibfnamefont{J.~D.} \bibnamefont{Urbina}}, \bibnamefont{and} \bibinfo{author}{\bibfnamefont{K.}~\bibnamefont{Richter}}, \bibinfo{journal}{Physical Review A} \textbf{\bibinfo{volume}{97}}, \bibinfo{pages}{061606} (\bibinfo{year}{2018})\relax
\mciteBstWouldAddEndPuncttrue
\mciteSetBstMidEndSepPunct{\mcitedefaultmidpunct}
{\mcitedefaultendpunct}{\mcitedefaultseppunct}\relax
\EndOfBibitem
\bibitem[{\citenamefont{Polkovnikov}(2003)}]{PhysRevA.68.053604}
\bibinfo{author}{\bibfnamefont{A.}~\bibnamefont{Polkovnikov}}, \bibinfo{journal}{Physical Review A} \textbf{\bibinfo{volume}{68}}, \bibinfo{pages}{053604} (\bibinfo{year}{2003})\relax
\mciteBstWouldAddEndPuncttrue
\mciteSetBstMidEndSepPunct{\mcitedefaultmidpunct}
{\mcitedefaultendpunct}{\mcitedefaultseppunct}\relax
\EndOfBibitem
\bibitem[{\citenamefont{Pappalardi et~al.}(2020)\citenamefont{Pappalardi, Polkovnikov, and Silva}}]{pappalardi2020quantum}
\bibinfo{author}{\bibfnamefont{S.}~\bibnamefont{Pappalardi}}, \bibinfo{author}{\bibfnamefont{A.}~\bibnamefont{Polkovnikov}}, \bibnamefont{and} \bibinfo{author}{\bibfnamefont{A.}~\bibnamefont{Silva}}, \bibinfo{journal}{SciPost Physics} \textbf{\bibinfo{volume}{9}}, \bibinfo{pages}{021} (\bibinfo{year}{2020})\relax
\mciteBstWouldAddEndPuncttrue
\mciteSetBstMidEndSepPunct{\mcitedefaultmidpunct}
{\mcitedefaultendpunct}{\mcitedefaultseppunct}\relax
\EndOfBibitem
\bibitem[{Note2()}]{Note2}
Note2, \bibinfo{note}{if necessary, systematic degeneracies due to symmetries are excluded by focusing on the eigenfunctions within a given symmetry-related subspace with fixed values of the corresponding quantum numbers}\relax
\mciteBstWouldAddEndPuncttrue
\mciteSetBstMidEndSepPunct{\mcitedefaultmidpunct}
{\mcitedefaultendpunct}{\mcitedefaultseppunct}\relax
\EndOfBibitem
\bibitem[{Note3()}]{Note3}
Note3, \bibinfo{note}{we assume $\eta $ to be large compared to the local mean level spacing which implies a large number of eigenstates inside the spectral window.}\relax
\mciteBstWouldAddEndPunctfalse
\mciteSetBstMidEndSepPunct{\mcitedefaultmidpunct}
{}{\mcitedefaultseppunct}\relax
\EndOfBibitem
\bibitem[{\citenamefont{Nakerst and Haque}(2021)}]{nakerst2021eigenstate}
\bibinfo{author}{\bibfnamefont{G.}~\bibnamefont{Nakerst}} \bibnamefont{and} \bibinfo{author}{\bibfnamefont{M.}~\bibnamefont{Haque}}, \bibinfo{journal}{Physical Review E} \textbf{\bibinfo{volume}{103}}, \bibinfo{pages}{042109} (\bibinfo{year}{2021})\relax
\mciteBstWouldAddEndPuncttrue
\mciteSetBstMidEndSepPunct{\mcitedefaultmidpunct}
{\mcitedefaultendpunct}{\mcitedefaultseppunct}\relax
\EndOfBibitem
\bibitem[{\citenamefont{Heller and Landry}(2007)}]{Heller_2007}
\bibinfo{author}{\bibfnamefont{E.~J.} \bibnamefont{Heller}} \bibnamefont{and} \bibinfo{author}{\bibfnamefont{B.~R.} \bibnamefont{Landry}}, \bibinfo{journal}{Journal of Physics A: Mathematical and Theoretical} \textbf{\bibinfo{volume}{40}}, \bibinfo{pages}{9259} (\bibinfo{year}{2007})\relax
\mciteBstWouldAddEndPuncttrue
\mciteSetBstMidEndSepPunct{\mcitedefaultmidpunct}
{\mcitedefaultendpunct}{\mcitedefaultseppunct}\relax
\EndOfBibitem
\bibitem[{Note4()}]{Note4}
Note4, \bibinfo{note}{knowledge of $\protect \hat {R}(E)$ provides all information required for the calculation of the microcanonical (mc) expectation value of any MB observable $\protect \hat {O}$, as $\langle \protect \hat {O} \rangle ^{(\protect \rm mc)}_{E}={\protect \rm Tr}\left [\protect \hat {R}(E)\protect \hat {O}\right ]$. Its exact determination is therefore a formidable task in general}\relax
\mciteBstWouldAddEndPuncttrue
\mciteSetBstMidEndSepPunct{\mcitedefaultmidpunct}
{\mcitedefaultendpunct}{\mcitedefaultseppunct}\relax
\EndOfBibitem
\bibitem[{\citenamefont{Fisher et~al.}(1989)\citenamefont{Fisher, Weichman, Grinstein, and Fisher}}]{PhysRevB.40.546}
\bibinfo{author}{\bibfnamefont{M.~P.~A.} \bibnamefont{Fisher}}, \bibinfo{author}{\bibfnamefont{P.~B.} \bibnamefont{Weichman}}, \bibinfo{author}{\bibfnamefont{G.}~\bibnamefont{Grinstein}}, \bibnamefont{and} \bibinfo{author}{\bibfnamefont{D.~S.} \bibnamefont{Fisher}}, \bibinfo{journal}{Physical Review B} \textbf{\bibinfo{volume}{40}}, \bibinfo{pages}{546} (\bibinfo{year}{1989})\relax
\mciteBstWouldAddEndPuncttrue
\mciteSetBstMidEndSepPunct{\mcitedefaultmidpunct}
{\mcitedefaultendpunct}{\mcitedefaultseppunct}\relax
\EndOfBibitem
\bibitem[{\citenamefont{Krutitsky}(2016)}]{krutitsky2016ultracold}
\bibinfo{author}{\bibfnamefont{K.~V.} \bibnamefont{Krutitsky}}, \bibinfo{journal}{Physics Reports} \textbf{\bibinfo{volume}{607}}, \bibinfo{pages}{1} (\bibinfo{year}{2016})\relax
\mciteBstWouldAddEndPuncttrue
\mciteSetBstMidEndSepPunct{\mcitedefaultmidpunct}
{\mcitedefaultendpunct}{\mcitedefaultseppunct}\relax
\EndOfBibitem
\bibitem[{\citenamefont{Bloch et~al.}(2008{\natexlab{b}})\citenamefont{Bloch, Dalibard, and Zwerger}}]{bloch2008many}
\bibinfo{author}{\bibfnamefont{I.}~\bibnamefont{Bloch}}, \bibinfo{author}{\bibfnamefont{J.}~\bibnamefont{Dalibard}}, \bibnamefont{and} \bibinfo{author}{\bibfnamefont{W.}~\bibnamefont{Zwerger}}, \bibinfo{journal}{Reviews of modern physics} \textbf{\bibinfo{volume}{80}}, \bibinfo{pages}{885} (\bibinfo{year}{2008}{\natexlab{b}})\relax
\mciteBstWouldAddEndPuncttrue
\mciteSetBstMidEndSepPunct{\mcitedefaultmidpunct}
{\mcitedefaultendpunct}{\mcitedefaultseppunct}\relax
\EndOfBibitem
\bibitem[{\citenamefont{Tomsovic}(2018)}]{Steve}
\bibinfo{author}{\bibfnamefont{S.}~\bibnamefont{Tomsovic}}, \bibinfo{journal}{Physical Review E} \textbf{\bibinfo{volume}{98}}, \bibinfo{pages}{023301} (\bibinfo{year}{2018})\relax
\mciteBstWouldAddEndPuncttrue
\mciteSetBstMidEndSepPunct{\mcitedefaultmidpunct}
{\mcitedefaultendpunct}{\mcitedefaultseppunct}\relax
\EndOfBibitem
\bibitem[{Note5()}]{Note5}
Note5, \bibinfo{note}{we do not address the issue of defining the Wigner transform on discrete spaces, as it is irrelevant in the semiclassical regime considered here.}\relax
\mciteBstWouldAddEndPunctfalse
\mciteSetBstMidEndSepPunct{\mcitedefaultmidpunct}
{}{\mcitedefaultseppunct}\relax
\EndOfBibitem
\bibitem[{\citenamefont{Voros}(1976)}]{voros1976semi}
\bibinfo{author}{\bibfnamefont{A.}~\bibnamefont{Voros}}, \bibinfo{journal}{Annales de l'Institut Henri Poincar\'e} \textbf{\bibinfo{volume}{24}} (\bibinfo{year}{1976})\relax
\mciteBstWouldAddEndPuncttrue
\mciteSetBstMidEndSepPunct{\mcitedefaultmidpunct}
{\mcitedefaultendpunct}{\mcitedefaultseppunct}\relax
\EndOfBibitem
\bibitem[{\citenamefont{Berry}(1977{\natexlab{b}})}]{berry1977semi}
\bibinfo{author}{\bibfnamefont{M.~V.} \bibnamefont{Berry}}, \bibinfo{journal}{Philosophical Transactions of the Royal Society of London. Series A, Mathematical and Physical Sciences} \textbf{\bibinfo{volume}{287}}, \bibinfo{pages}{237} (\bibinfo{year}{1977}{\natexlab{b}})\relax
\mciteBstWouldAddEndPuncttrue
\mciteSetBstMidEndSepPunct{\mcitedefaultmidpunct}
{\mcitedefaultendpunct}{\mcitedefaultseppunct}\relax
\EndOfBibitem
\bibitem[{Note6()}]{Note6}
Note6, \bibinfo{note}{see Supplemental Material at [url], which further includes \cite {Fischer_2013} for additional information about the derivation of the Eq.(\ref {eq:Rsc}) and \cite {beugeling2018statistical} for an in depth discussion about statistical properties of eigenstate amplitudes in complex quantum systems.}\relax
\mciteBstWouldAddEndPunctfalse
\mciteSetBstMidEndSepPunct{\mcitedefaultmidpunct}
{}{\mcitedefaultseppunct}\relax
\EndOfBibitem
\bibitem[{\citenamefont{Atas and Bogomolny}(2017)}]{atas2017quantum}
\bibinfo{author}{\bibfnamefont{Y.}~\bibnamefont{Atas}} \bibnamefont{and} \bibinfo{author}{\bibfnamefont{E.}~\bibnamefont{Bogomolny}}, \bibinfo{journal}{Journal of Physics A: Mathematical and Theoretical} \textbf{\bibinfo{volume}{50}}, \bibinfo{pages}{385102} (\bibinfo{year}{2017})\relax
\mciteBstWouldAddEndPuncttrue
\mciteSetBstMidEndSepPunct{\mcitedefaultmidpunct}
{\mcitedefaultendpunct}{\mcitedefaultseppunct}\relax
\EndOfBibitem
\bibitem[{\citenamefont{Mirlin}(2000)}]{MIRLIN2000}
\bibinfo{author}{\bibfnamefont{A.~D.} \bibnamefont{Mirlin}}, \bibinfo{journal}{Physics Reports} \textbf{\bibinfo{volume}{326}}, \bibinfo{pages}{259} (\bibinfo{year}{2000})\relax
\mciteBstWouldAddEndPuncttrue
\mciteSetBstMidEndSepPunct{\mcitedefaultmidpunct}
{\mcitedefaultendpunct}{\mcitedefaultseppunct}\relax
\EndOfBibitem
\bibitem[{Note7()}]{Note7}
Note7, \bibinfo{note}{similar results are obtained for any seed state under the conditions discussed below}\relax
\mciteBstWouldAddEndPuncttrue
\mciteSetBstMidEndSepPunct{\mcitedefaultmidpunct}
{\mcitedefaultendpunct}{\mcitedefaultseppunct}\relax
\EndOfBibitem
\bibitem[{\citenamefont{Khaymovich et~al.}(2019)\citenamefont{Khaymovich, Haque, and McClarty}}]{Haque2019}
\bibinfo{author}{\bibfnamefont{I.~M.} \bibnamefont{Khaymovich}}, \bibinfo{author}{\bibfnamefont{M.}~\bibnamefont{Haque}}, \bibnamefont{and} \bibinfo{author}{\bibfnamefont{P.~A.} \bibnamefont{McClarty}}, \bibinfo{journal}{Phys. Rev. Lett.} \textbf{\bibinfo{volume}{122}}, \bibinfo{pages}{070601} (\bibinfo{year}{2019})\relax
\mciteBstWouldAddEndPuncttrue
\mciteSetBstMidEndSepPunct{\mcitedefaultmidpunct}
{\mcitedefaultendpunct}{\mcitedefaultseppunct}\relax
\EndOfBibitem
\bibitem[{\citenamefont{Fischer et~al.}(2013)\citenamefont{Fischer, Gneiting, and Hornberger}}]{Fischer_2013}
\bibinfo{author}{\bibfnamefont{T.}~\bibnamefont{Fischer}}, \bibinfo{author}{\bibfnamefont{C.}~\bibnamefont{Gneiting}}, \bibnamefont{and} \bibinfo{author}{\bibfnamefont{K.}~\bibnamefont{Hornberger}}, \bibinfo{journal}{New Journal of Physics} \textbf{\bibinfo{volume}{15}}, \bibinfo{pages}{063004} (\bibinfo{year}{2013})\relax
\mciteBstWouldAddEndPuncttrue
\mciteSetBstMidEndSepPunct{\mcitedefaultmidpunct}
{\mcitedefaultendpunct}{\mcitedefaultseppunct}\relax
\EndOfBibitem
\bibitem[{\citenamefont{Beugeling et~al.}(2018)\citenamefont{Beugeling, B{\"a}cker, Moessner, and Haque}}]{beugeling2018statistical}
\bibinfo{author}{\bibfnamefont{W.}~\bibnamefont{Beugeling}}, \bibinfo{author}{\bibfnamefont{A.}~\bibnamefont{B{\"a}cker}}, \bibinfo{author}{\bibfnamefont{R.}~\bibnamefont{Moessner}}, \bibnamefont{and} \bibinfo{author}{\bibfnamefont{M.}~\bibnamefont{Haque}}, \bibinfo{journal}{Physical Review E} \textbf{\bibinfo{volume}{98}}, \bibinfo{pages}{022204} (\bibinfo{year}{2018})\relax
\mciteBstWouldAddEndPuncttrue
\mciteSetBstMidEndSepPunct{\mcitedefaultmidpunct}
{\mcitedefaultendpunct}{\mcitedefaultseppunct}\relax
\EndOfBibitem
\end{mcitethebibliography}
\end{document}